\documentclass{aa}  
\usepackage{lastpage}

\usepackage{graphicx}

\usepackage{txfonts}

\usepackage{booktabs, makecell}
\usepackage{dirtytalk}
\usepackage{tabularx}
\usepackage{amsmath}
\usepackage{upgreek} 

\usepackage{float}
\usepackage{placeins}

\usepackage[T1]{fontenc}
\usepackage{courier}
\usepackage[colorlinks=true,
            linkcolor=blue,
            urlcolor=blue,
            citecolor=blue]{hyperref}

\begin{document}

   \title{Disentangling Auroral, Cloud and Magnetic Spot Driven Variability in Three Early L-dwarfs with HST/WFC3}

   \author{C. O'Toole
          \inst{1}
          \and J. M. Vos \inst{1}
          \and E. N. Nasedkin \inst{1}
          \and J. S. Pineda \inst{2}
          \and M. M Kao \inst{3}
          \and Y. Zhou \inst{4}
          \and M. Schrader \inst{1}
          \and A. M. McCarthy \inst{5,} \inst{1}
          }

   \institute{School of Physics, Trinity College Dublin, The University of Dublin, Dublin 2, Ireland; \email{cotoole3@tcd.ie}              
        \and University of Colorado Boulder, Laboratory for Atmospheric and Space Physics, 3665 Discovery Drive, Boulder, CO 80303, USA
        \and Lowell Observatory, 1400 W Mars Hill Road, Flagstaff, AZ 86001, USA
        \and Department of Astronomy, University of Virginia, 530 McCormick Road, Charlottesville, VA 22904, USA
        \and Department of Astronomy \& The Institute for Astrophysical Research, Boston University, 725 Commonwealth Avenue, Boston, MA 02215, USA\\
            }

   \date{Received December 04, 2025; accepted March 24, 2026}

  \abstract 
    {Variability monitoring provides an unparalleled insight into the atmospheric processes of brown dwarfs and directly imaged exoplanets. Inhomogeneous clouds, aurora{e} and magnetic spots have all been postulated as potential drivers of variability. While objects at the L/T transition have had their variability studied extensively, the variability of early L-dwarfs remain{s} an understudied region of the parameter space. We use observations from the Hubble Space Telescope in the near-infrared, using WFC3/G141  to disentangle the drivers of variability in three known variable early L-dwarfs: 2MASS J1721039+334415, 2MASS J00361617+1821104 and 2MASS J19064801+4011089.
    We find that all three objects exhibit significant variability at all wavelengths, with white-light amplitudes of 0.53-1.41 $\%$. We find that their colour variations are  brighter and bluer compared to later spectral types, except for 2MASS J19064801+4011089 which exhibits largely grey variations.
    We report a new period for 2MASS J1721039+334415, of $4.9^{+0.4}_{-0.2}$ hours. We find evidence of long term light curve stability in each object, which may indicate the presence of long lived features on their surfaces.
    We create a flexible modelling framework to model three potential drivers of variability: clouds, aurorae and magnetic spots. We fit our models to the spectral variability amplitude from 1.1-1.67 $\upmu$m of each object. We find that changing cloud properties or magnetic spots are the most likely drivers of variability in each object.
    Auroral models do not reproduce the variability within the HST wavelengths, however future observations at longer wavelengths that probe higher in the atmosphere may be more sensitive to auroral effects. This work provides a foundation for future variability studies of early L-dwarfs and directly imaged exoplanets to disentangle auroral, cloud and magnetic spot driven variability. }


   \maketitle
%
\begin{table*}
\centering
\renewcommand{\arraystretch}{1.1}
\caption{Fundamental parameters and measured variability parameters for each object in our sample.} 

\begin{tabular}{lll|ll|ll}
\toprule
\textbf{Object} & \textbf{2M1721+33} & \textbf{Ref} & \textbf{2M0036+18} & \textbf{Ref} & \textbf{2M1906+40} & \textbf{Ref} \\
\midrule

\textbf{Spectral Type} & L3  & (K03) & L3.5 &  (R00) & L1  & (G11)\\

\textbf{T$_{eff}$ (K)} & 1802 $\pm$ 174 & (S23) & 1869 $\pm$ 64  &  (F15) & 2201 $\pm$ 202 & (S23) \\

\textbf{Log (g)} &5.15 $\pm$ 0.43  & (S23) & 5.21$\pm$0.17  &  (F15)  & 5.23 $\pm$ 0.39  &  (S23) \\

\textbf{Literature Period (Hr)} & $2.6\pm0.1$ & (M15)  & $3.080\pm0.001$  &  (C16) & ${8.88364}\pm5\times10^{-5}$ &  (G15) \\

\textbf{GP Period (Hr)} & $4.93^{+0.36}_{-0.22}$ & (TW) & $3.13^{+0.09}_{-0.08}$ & (TW) & $19.4^{+46.5}_{-10.5}$  & (TW) \\

\textbf{Light Curve Amplitude (\%)} & 
$0.64^{+0.03}_{-0.02}$  & (TW) & $1.41^{+0.16}_{-0.02}$ (obs 1)  & (TW) & $0.53^{+0.03}_{-0.01}$ & (TW) \\[2pt]

 &  &  & $1.36^{+0.11}_{-0.01}$ (obs 2) & (TW) &  &  \\[2pt]

\bottomrule
\end{tabular}
\tablebib{K03= \cite{SpecType_1721_2003},  R00 = \cite{Reid2000},
G11 = \cite{Gizis_2011},
M15 = \cite{metchev2015weather},
F15 = \cite{filipazzo2015}, G15 = \cite{gizis2015kepler}, C16 = \cite{croll2016long}, S23 = \cite{Sanghi2023}, TW = This Work.}
\vspace{2mm}
\label{tab: periods}
\end{table*}

\section{Introduction}
Brown dwarfs are often regarded as analogues to directly imaged giant exoplanets, as they share many similar properties such as  radii, effective temperature, thermal profiles, surface gravity and chemistry \citep[{e.g.}][]{Faherty2016, Liu_2016}. 
The majority of brown dwarf and brown dwarf companion observations are not limited by the extreme star-to-planet contrasts of exoplanet observations which enables more detailed analyses, particularly for time-resolved studies \citep[{e.g.}][]{Apai_2013, Zhou2020VHS1256}.

Variability monitoring allows for detailed exploration of extrasolar atmospheres and the dynamical processes that shape them via measurements of an object’s luminosity as it rotates.
Variability studies with spaced-based facilities such as the Hubble Space Telescope (HST) \citep[{e.g.}][]{Apai_2013, yang2015, bowler2020strong}, the Spitzer Space Telescope \citep[{e.g.}][]{metchev2015weather, Vos2018, Vos2022spin} and more recently the James Webb Space Telescope (JWST) \citep[{e.g.}][]{Biller_2024,Chen2025, McCarthy2025_JWST_SIMP, nasedkin2025}, have provided unparalleled insights into the dynamics of brown dwarf atmospheres.

Photometric variability monitoring across the optical, near-infrared (near-IR) and mid-infrared (mid-IR) has  shown that dramatic variability is common among brown dwarfs \citep[{e.g.}][]{Radigan2014, metchev2015weather,Vos2022spin}. This variability has been attributed to a number of different processes, from patchy cloud cover \citep[{e.g.}][]{Radigan2014, Vos2023PatchyAnalogs} to auroral processes \citep[]{Hallinan2015}, and magnetic spots \citep[]{gizis2015kepler, croll2016long}. However, disentangling the processes that drive variability requires spectroscopic monitoring.

Spectroscopic variability monitoring probes different pressures of the atmosphere simultaneously \citep[{e.g.}][]{Biller_2024, McCarthy2025_JWST_SIMP,Chen2025}, and provides a method to probe the mechanisms that drive variability. HST has paved the way for these studies, allowing us to peer into the vertical structure of these atmospheres \citep[]{buenzli2012, Apai_2013}. Modelling how the variability changes with wavelength (the spectral variability amplitude), is key to understanding what processes are driving the variability. Cloud-driven variability models have been the primary method to reproduce the spectral variability amplitude for L and T dwarfs \citep[{e.g.}][]{Apai_2013, yang2015, bowler2020strong}. HST has shown that the spectral variability amplitude changes depending on the spectral type of the object. 
This is due to the fact that different spectral bands probe different atmospheric pressures.
This is particularly evident within the water band feature: for L-dwarfs, the spectral variability amplitude is characterised by high-altitude hazes and is consistent both in and out of the water feature while L/T transition objects have lower variability in the water feature.
At the L/T transition, clouds are typically located at lower altitudes (higher pressures) compared to L-dwarfs, resulting in lower variability in the water band \citep[{e.g.}][]{yang2015, Lew2020CloudDwarf, bowler2020strong}. {Planetary scale waves have also been proposed as a mechanism of cloud driven variability on brown dwarfs \citep[e.g.][]{Apai2017,Zhou2022RoaringAtmosphere,Fuda2024}, and have been successful at constraining the driver of variability for observations spanning hundreds of rotations \citep{Fuda2024}.} 

\begin{figure*}
    \centering
    \includegraphics[width=1\textwidth]{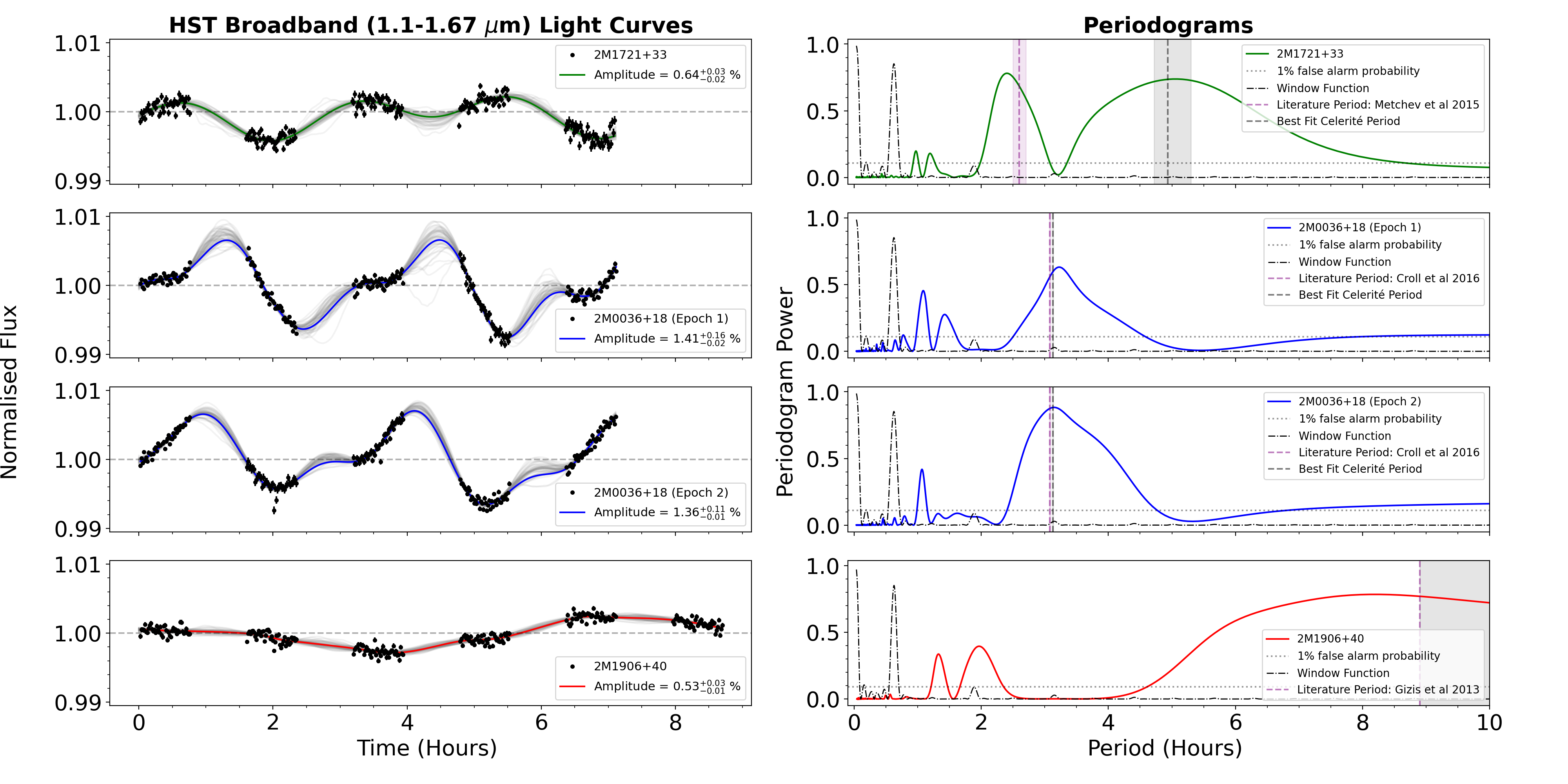}
    \caption{\textbf{Left:} The broadband G141 filter light curves of each of our observations. A sample of 100 posterior fits of each light curve are shown in grey, while the best fit light curve is shown in colour for each observation. All four light curves are variable. \textbf{Right:} The periodogram for each light curve. {The periodogram for the window function of HST is shown by the dash-dot black line in each plot.}
    The period of each object from the literature is shown by the purple vertical line in the periodogram, while the period output from our {\tt celerité2} models are shown by the black vertical lines. The uncertainties for the period values from both the literature and this work are represented by the shaded regions in purple and black respectively. Our model periods agree with the literature for each observation except 2M1721+33, where we estimate a longer period of $4.93^{+0.36}_{-0.22}$ hours. 
    }
    \label{fig:LCs+period}
\end{figure*}

The variability in early L-dwarfs, $<$ L4, has not been studied as extensively as their cooler counterparts at the L/T transition. Early L-dwarfs can provide an understanding of how atmospheres transition across the stellar-substellar boundary, as they are at the M/L dwarf transition, where there is an overlap between the coolest M dwarfs and hottest L-dwarfs.
Variability in this regime may be caused by a number of mechanisms, such as clouds, aurorae or magnetic spots, \citep[{e.g.}][]{Apai_2013, Hallinan2015, Barnes2015,Kao2016AURORALREGIME,zuckerman2026}. While cloud driven variability has been extensively investigated with HST, \citep[{e.g.}][]{Apai_2013,yang2015,bowler2020strong}, auroral or magnetic spot driven variability have not.

Variability driven by the presence of magnetic spots has been hypothesized for early L-dwarfs, based on observations with ground based telescopes and the Kepler Space Telescope \citep{Lane2007,Gizis2013, croll2016long}.
Magnetic spot driven variability has been observed on M dwarfs \citep[{e.g.}][]{ Rockenfeller2006, Lane2007, Mclean2011,Rackham2018}.
{As early L-dwarfs straddle the stellar/substellar boundary, it is plausible that they may also host magnetic spots that can drive variability. Late M-dwarfs (M7-M9) commonly exhibit magnetic spot driven variability \citep[{e.g.}][]{Lane2007, Mclean2011,Rackham2018}. }

\cite{L5_flares_2020}, detected white light flares on 9 L dwarfs, spanning spectral types from L0-L5 using observations from the K2 mission. 
As a result of the K2 observations spanning 4402 days, they had a long baseline to observe flaring activity. 
Based on results from \citet[]{deluca2020,Baylor2011}, \citet{L5_flares_2020}, find that flares in a range of stars, from spectral type G to L, share a number of properties despite the difference in their internal structures. They estimate that a 10$^{33}$ erg superflare occurs on early and mid-L dwarfs approximately once every 2.4 years. This low flaring frequency highlights why so few L dwarf flares are found in the literature as typical observing campaigns of brown dwarfs with facilities such as JWST, HST and Spitzer are on the order of 10s-100s of hours, and thus are highly unlikely to observe such a flare. Magnetic spots that produce these flares therefore may be ubiquitous in the early to mid-L dwarf regime, and may drive variability on these objects.
Therefore early L-dwarfs with strong magnetic fields {are among} the coolest objects where magnetic spot driven variability may occur.
However, magnetic spot driven variability of brown dwarfs have yet to be {explored} in the context of HST observations.

Thermochemical instabilities driven by diabatic processes have also been proposed as a potential driver of variability, \citep[]{Tremblin2019,Tremblin2020,oliveros2026}. {These mechanisms are typically associated with the L/T transition where CO is converted into CH$_{4}$ \citep[]{Tremblin2019,Tremblin2020}. Therefore, we can rule this method out as it is dependent on the presence of CH$_{4}$, which is not expected at these spectral types.}

Using spectroscopic variability observations of three early L dwarfs obtained with HST/WFC3, we seek to disentangle their drivers of variability by analysing their spectral variability amplitude. This study is structured as follows.
In Section \ref{sec: sample}, we outline the properties of the sample.
In Section \ref{sec:observations} we describe the HST observations and data reduction, including the light curve extraction, Gaussian process modelling and periodogram analysis. We illustrate the colour modulation of each object in Section \ref{sec:colour modulation}, and highlight these objects as a complementary parameter space to previously studied brown dwarfs. In Section \ref{sec:spectral variability amplitude}, we present the wavelength dependence of the variability amplitude for each object and we outline our modelling framework for multiple drivers of variability. We then discuss the results of our modelling in Section \ref{sec: discussion}, evaluate what are the most likely drivers of variability in our objects and compare them to their light curves in the literature. The conclusions are summarized in Section \ref{sec:conclusions}.

\section{Sample properties}\label{sec: sample}

We chose our 3 L-dwarf targets to characterize and test near-IR signatures of clouds, aurorae and magnetic fields. 2MASS J00361617+1821104 (hereafter 2M0036+18) was selected as a known auroral object likely to possess strong magnetic fields, which we contrast with 2MASS J1721039+334415 (hereafter 2M1721+33) which has weak magnetic activity and known cloud behaviours. Against these two benchmarks we compare data for our third object, 2MASS J19064801+4011089 (hereafter 2M1906+40) {with observed variability over 2200 rotations, which may be due to long lived surface features}. We proceed with the characteristics of individual objects as follows:

2M0036+18, is an aurorally emitting L3.5 dwarf. We show its fundamental parameters in  Table \ref{tab: periods}. It emits coherent radio emission characteristic of aurora, including consistent periodic radio pulses that are highly circularly polarized with high brightness temperatures, {$P_{Radio}=3.08\pm0.05$ hours} \citep{Hallinan2008}. Its radio aurora has also been consistent over $>$10 years \citep{Berger2002_Aurora,Hallinan2008}. 
2M0036+18 has shown photometric variability in the optical, near-IR and mid-IR \citep[]{Harding2013,metchev2015weather, croll2016long}. This makes it the ideal candidate to search for aurorally driven variability in the infrared. 2M0036+18 has
an {infrared photometric} period of 3.080 $\pm0.001$ hours \cite{croll2016long} and 
has previously had near-IR variability monitoring with the Spitzer Space Telescope \cite{metchev2015weather} at 3.6 and 4.5 $\mu$m.

2M1721+33, is an L3 dwarf whose variability is thought to be driven by clouds {\cite{metchev2015weather}}. Its fundamental parameters are shown in Table \ref{tab: periods}. It is magnetically inactive and thus unlikely to host aurorae or magnetic spots \citep{Ldwarf_mag_spot2015,Richey_Yowell2020}.  
2M1721+33 is an ideal candidate to act as a control for cloud-driven variability, against which we can contrast the variability of our other targets. This is strengthened by the fact that 2M1721+33 has a silicate feature at 10 $\mu$m, indicative of a cloudy atmosphere \citep{Suarez2022}.
2M1721+33 had mid-IR variability monitoring with the Spitzer Space Telescope at 3.6 and 4.5 $\mu$m, with a measured {infrared photometric} period of 2.6 $\pm0.1$ hours \citep{metchev2015weather}.

{2M1906+40 is an L1 dwarf whose variability {has been attributed to} long-lived surface features such as magnetic spots or an inhomogeneous cloud deck \citep[]{Gizis2013, gizis2015kepler}.}
Its fundamental parameters are shown in Table \ref{tab: periods}. Long-term observations with the Kepler Space Telescope revealed that its light curve remained stable over 2200 rotations \citep{gizis2015kepler}. This {light curve stability} is unusual compared to light curve evolution driven by clouds on rotational timescales, \citep[]{Karaldi2016,Apai2017,Vos2018}. {\citet{Gizis2013} suggest that the variability may be due to magnetic spots, but} \citet{gizis2015kepler} favour a stable cloud spot model to reproduce the observed variability.
However, neither auroral driven {nor magnetic spot driven} variability has been ruled out.
The stable nature of 2M1906+40's light curve suggests that there is a long lived surface feature driving its variability. 
\citet{gizis2015kepler} present mid-IR monitoring at 3.6 and 4.5 $\mu$m with the Spitzer Space Telescope. \citep{Gizis2013,gizis2015kepler} measure an {optical photometric} period of {8.88364} $\pm 5 \times 10^{-5}$ hours using its Kepler light curve.
By analysing {our} sample as a whole, we can place this object in context by comparing it with others that show evidence for cloud-driven variability and auroral activity.

\begin{table*}[t]
\centering

\caption{Observation details for the three targets, from HST GO 15924. }
{\begin{tabular}{lccccc}

\toprule

\makecell{\thead{\textbf{ Object}}} & \thead{\textbf{Orbits}} & \thead{\textbf{Observation Dates (UTC)}} & \thead{\textbf{Observation Times (UTC)}} & \thead{\textbf{Total Exposures}} & \thead{\textbf{Time per Exposure (s)}} \\

\midrule

2M1906+40 & 6 & 07-08  December  2020 & 21:37 - 06:19 &  258 & 63  \\
2M1721+33 & 5 & 16-17  October  2020 & 23:56 - 07:02 &  215 & 63  \\
2M0036+18 & 5 & 26  July  2020 & 12:43 - 19:50 & 215 & 63 \\
2M0036+18 & 5 & 07  December  2021 & 00:33 - 07:40 &  215 & 63 
\\

\bottomrule
\end{tabular}}

\vspace{2mm} 
\label{tab: observations}
\end{table*}

\section{Observations and data reduction}\label{sec:observations}
\subsection{Observations}
Each target was observed simultaneously with the HST/WFC3 instrument and the VLA, (PID  \#15924, PI: Vos).
This provided multi-wavelength coverage across both the near-IR and the radio. Analysis of the VLA observations will be reported in a future publication.
HST's WFC3/IR camera was used to observe the three targets using the G141 grism over a wavelength range of $1.1 - 1.67~\upmu$m. These observations followed the observing strategy demonstrated by numerous successful examples of brown dwarf HST/WFC3 grism monitoring, \citep[{e.g.}][]{buenzli2012,Apai_2013, yang2015}.

The G141 grism has a {resolving power $\approx$ 130} over the range of $1.1-1.67~\upmu$m which provides access to the $1.4~\upmu$m water feature that is inaccessible from the ground. 
A direct image was taken at the start of each orbit using the F127M filter to determine the location of the target on the detector and to provide the G141 grism spectrum’s wavelength zero point.
Every image was captured in a 256 x 256 pixel sub-array and each target was observed in staring-mode. 
Each image had an exposure time of 63 seconds, with 43 exposures per orbit. Further details of the observations are summarised in Table \ref{tab: observations}.

We observed 2M0036+18 and 2M1721+33 for five consecutive orbits in order to obtain full phase coverage. 
2M1906+40 was observed for six consecutive orbits to obtain $\sim$ 8.7 hours of observations, slightly less than its reported period of {8.88364}  $\pm 5 \times 10^{-5}$ hours, \cite{gizis2015kepler}.
We obtained two datasets for 2M0036+18 as the VLA was not available at the time of the first observation. Therefore, we conducted a repeat observation in December 2021 simultaneously with the VLA (see Table \ref{tab: observations}).

\begin{figure}
    \centering
    \includegraphics[width=\columnwidth]{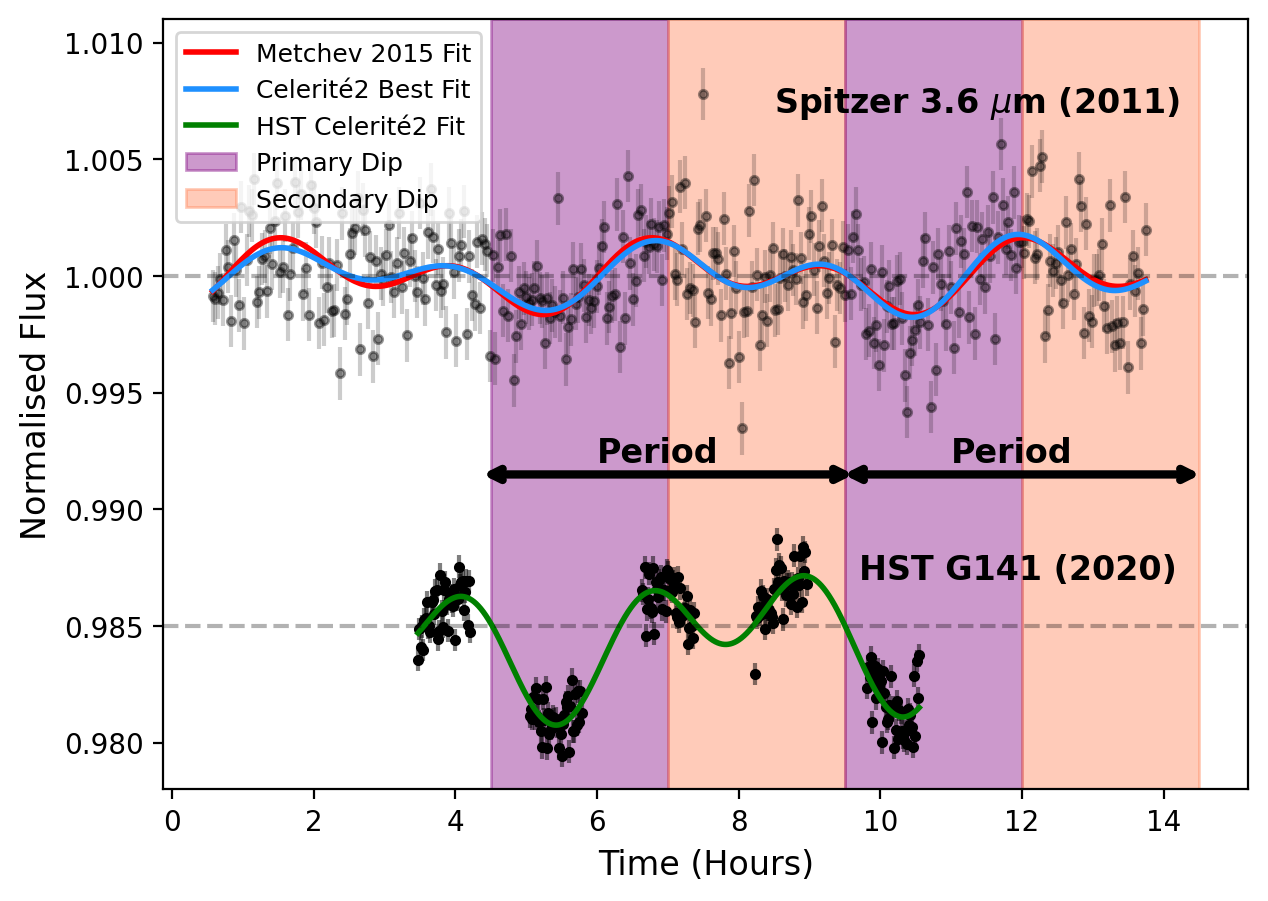}
    \caption{Long term stability of 2M1721+33 across two different wavebands. On top is the \cite{metchev2015weather} Spitzer 3.6~$\upmu$m light curve with both their fit in red and our Gaussian process {\tt celerité2} fit in blue applied to the Spitzer and HST light curves independently. Both models fit the data very similarly. On the bottom is the HST {white} light curve from our work with our Gaussian process {\tt celerité2} over-plotted (green). Despite these two light curves being observed over 9 years apart and at different wavelengths, there is a common shape to both light curves. They both exhibit a primary dip, followed by a secondary dip in each period. This may suggest long term stability in 2M1721+33's light curve. }

    \label{fig:Spitzer LC}
\end{figure}

\begin{figure}[h]
    \centering
    \includegraphics[width= \columnwidth]{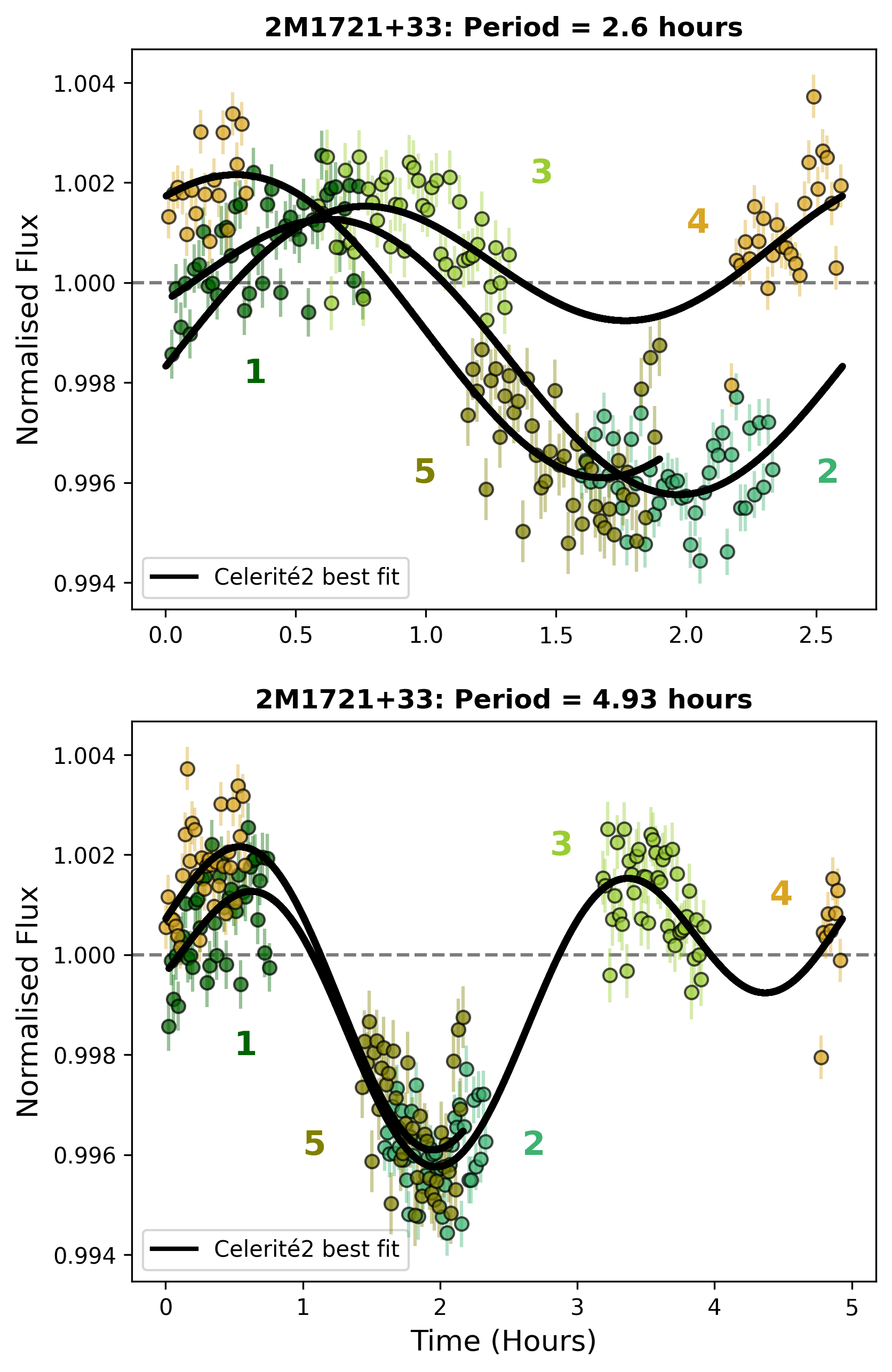}
    \caption{{ \textbf{Top:} Phase folded white light curve of 2M1721+33 using the the 2.6$\pm$0.1 hour period reported by \cite{metchev2015weather}. The photometric points from each HST orbit are colour-coded and the best fit {\tt celerité2} model is plotted in black. Using this period for phase folding does not give a clean light curve. \textbf{Bottom:} Phase folded white light curve according to the period of $4.9^{+0.4}_{-0.2}$ hours from this work. While this observation does not cover more than 1.5 rotations, the phase folded regions (orbits 1 and 4, and orbits 2 and 5) overlap with each other in both the HST data and the {\tt celerité2} model. This cleaner phase folded light curve, along with Figure \ref{fig:Spitzer LC} motivate our reasons to adopt this new period for 2M1721+33.
       }}
    \label{fig: 1721_phase_fold}
\end{figure}

\begin{figure}[h]
    \centering
    \includegraphics[width= \columnwidth]{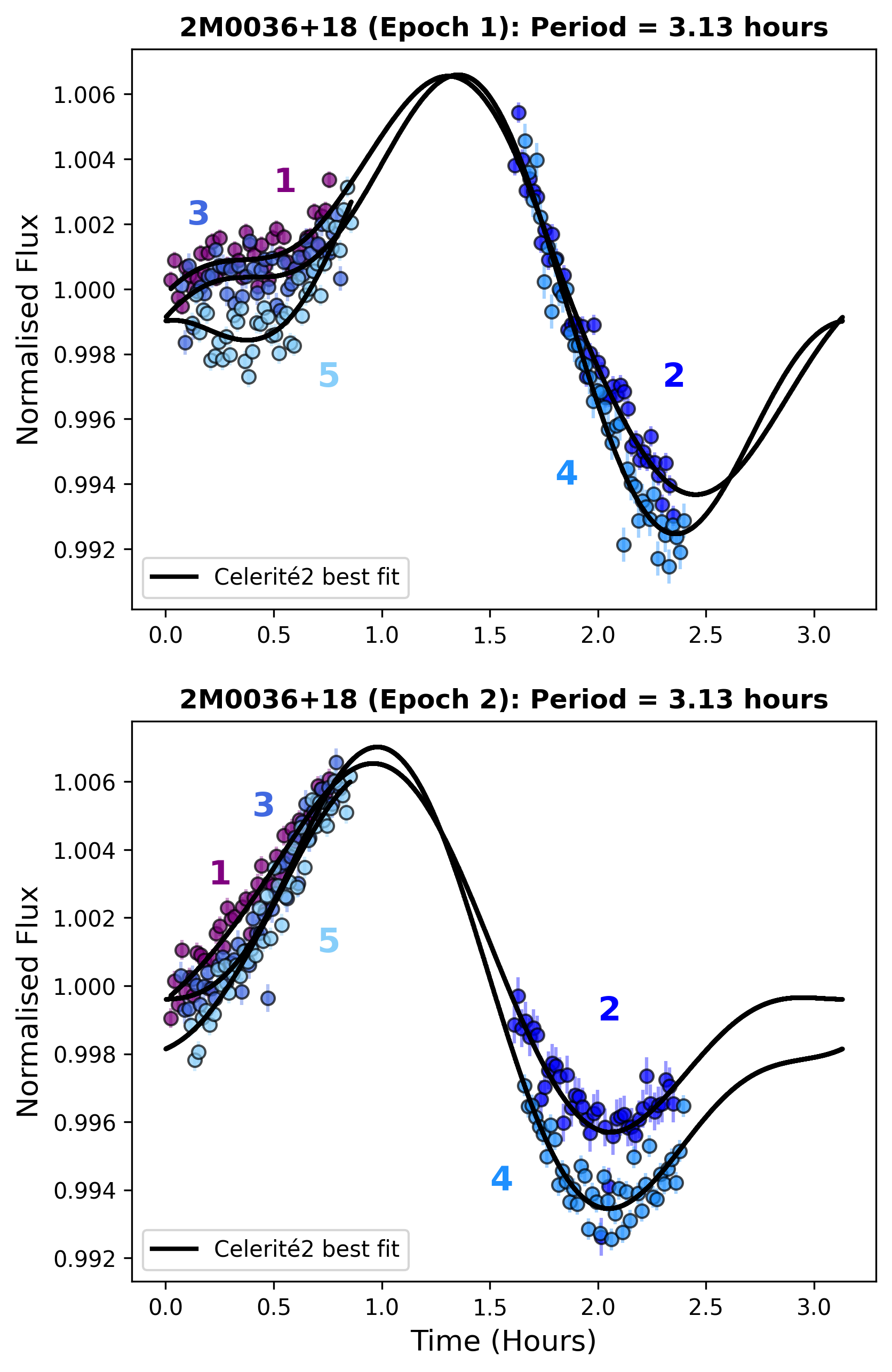}
    \caption{{\textbf{Top}: The phase folded white light curve of epoch 1 of 2M0036+18, using the period from this work ($3.13^{+0.9}_{-0.8}$ hours). The best fit {\tt celerité2} model folds neatly as does the HST white light curve. \textbf{Bottom:} The phase folded white light curve for epoch 2 of 2M0036+18 using the same period. As with the first epoch, both the best fit {\tt celerité2} model and the HST white light curve overlap neatly. Despite both epochs covering slightly different phases of the light curve, the same period works well for both.  }}
    \label{fig: 0036_phase_fold}
\end{figure}

\subsection{Data reduction}
In order to begin the spectral extraction process for our targets, the \textit{flt.fits} files -- produced by the {\tt calwf3} pipeline -- were downloaded from the Barbara A. Mikulski Archive for Space Telescopes (MAST). We used custom python scripts and \textit{aXe} \citep{kummel2009slitless} to reduce the data following best practices for HST/WFC3 \citep{buenzli2012,Apai_2013, yang2015}.

\begin{figure*}[t]
    \centering
    \includegraphics[width=0.97\textwidth]{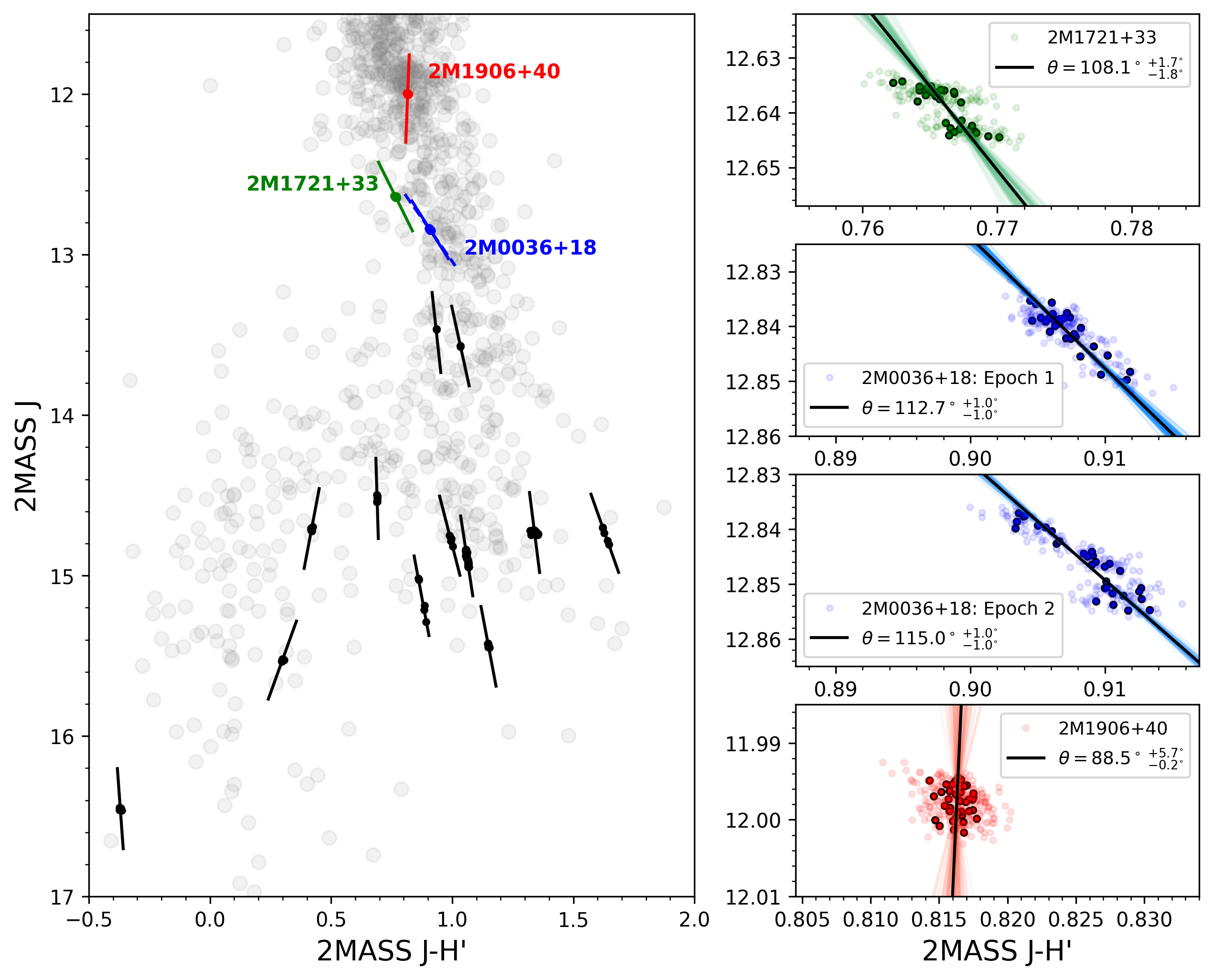}
    \caption{\textbf{Left:} Colour modulation of 2M1721+33 (green), 2M0036+18 (blue) and 2M1906+40 (red) compared with the sample from \cite{Lew_colour} (black circles and lines). The grey circles represent brown dwarfs with known parallaxes across spectral types, taken from the UltracoolSheet \cite{ultracool_sheet}. The slopes of 2M1721+33 and 2M0036+18 in particular are shallower in comparison  to the rest of the sample, showing that as these objects rotate, they are becoming brighter and bluer.
    \textbf{Right:} The 2MASS J vs 2MASS J-H' magnitudes for each observation ({dark circles are binned down to a 441 second cadence}) along with the individual fits for each brown dwarf using orthogonal distance regression (ODR), with their $\theta$ values and associated uncertainties. The colour modulation plot for 2M1721+33  appears as nearly two individual regions, however, this is due to a lack of phase coverage as a result of HST observations. The two observations of 2M0036+18 are plotted separately and they share a very similar slope, despite observations occurring 16 months apart. }
    \label{fig:colour mod}
\end{figure*}

We begin by identifying bad pixels by accessing the \say{Data Quality} extension from the \textit{flt.fits} files. These arrays use flag values of 4, 16, 32 and 256 to identify bad pixels, hot pixels, unstable response pixels and full-well saturation pixels respectively.
We correct these pixels using bi-linear interpolation of the four neighbouring pixels in the x and y directions.

Background sky subtraction was undertaken in the same manner as that of \cite{brammer2015source}. This considers both the zodiacal light background, i.e. the reflected solar continuum, and the spectral variation of the 1.083$\upmu$m emission line in the Earth's upper atmosphere, which varies in intensity over the course of an orbit. This algorithm is described fully in Appendix 6 of \cite{brammer2015source}.
Then, pixels that were flagged as cosmic rays were masked from the \textit{flt.fits} files by interpolating over the neighbouring pixels. 
Aperture extraction widths of 5.65, 5.50, 5.50, and 5.60 pixels in diameter were used for 2M1906+40, 2M1721+33, 2M0036+18 (observation 1) and 2M0036+18 (observation 2), respectively. These widths were chosen as they provided the highest signal to noise in our spectra for each observation.
We encountered wavelength calibration issues in the spectra of 2M1906+40. The assigned wavelengths were offset every second orbit, producing an artificial flux increase in the  fifth orbit of the light curve. To correct this, we fixed the wavelength reference for each orbit to the position measured in the F127M image from the first orbit.

\subsection{Light curve creation}\label{sec: light curves}

In order to analyse the variability of our objects, we created light curves. 
We applied $\sigma$-clipping to the spectra of each object to remove any outliers, discarding any points $>3\sigma$ away from the median spectrum of each observation.
Using our spectra, we produced light curves with the {\tt species} \citep{species} package for synthetic photometry, with the HST G141 filter.

We noted a clear ramp effect in our resulting light curves. The ramp effect manifests as a \say{tail} at the start of each orbit and is most pronounced in the first orbit of each observation. This tail starts off dimmer before rapidly increasing in brightness. This effect is systematic of HST/WFC3 IR time series observations \citep{buenzli2012,Apai_2013,yang2015}.
We used the Ramp Effect Charge Trapping Eliminator \citep[{\tt RECTE};][]{Zhou2017} to correct for the ramp effect.
{\tt RECTE} models the systematic ramp effect using a physically motivated method that takes into account electron charge trapping and the fast and slow lifetimes of these trap populations. Charge carriers generated by incident photons can become trapped as they diffuse across the depletion region of the photodiode. At the end of an exposure, the trapped charge carriers remain in the depletion region. This can lead to the ramp-shaped light curve in time series observations.
The free parameters of the model are the initial fast and slow trap populations, $E_{f,0}, E_{s,0}$, and their changes between orbits, $\Delta E_{f,0}, \Delta E_{s,0}$.
We constrained these parameters by fitting the RECTE model to the broadband G141 light curve, performing the fit separately for each observation. The final light curves for each observation are shown in Figure \ref{fig:LCs+period}.

\begin{figure*}[t]
    \centering
    \includegraphics[width= \textwidth]{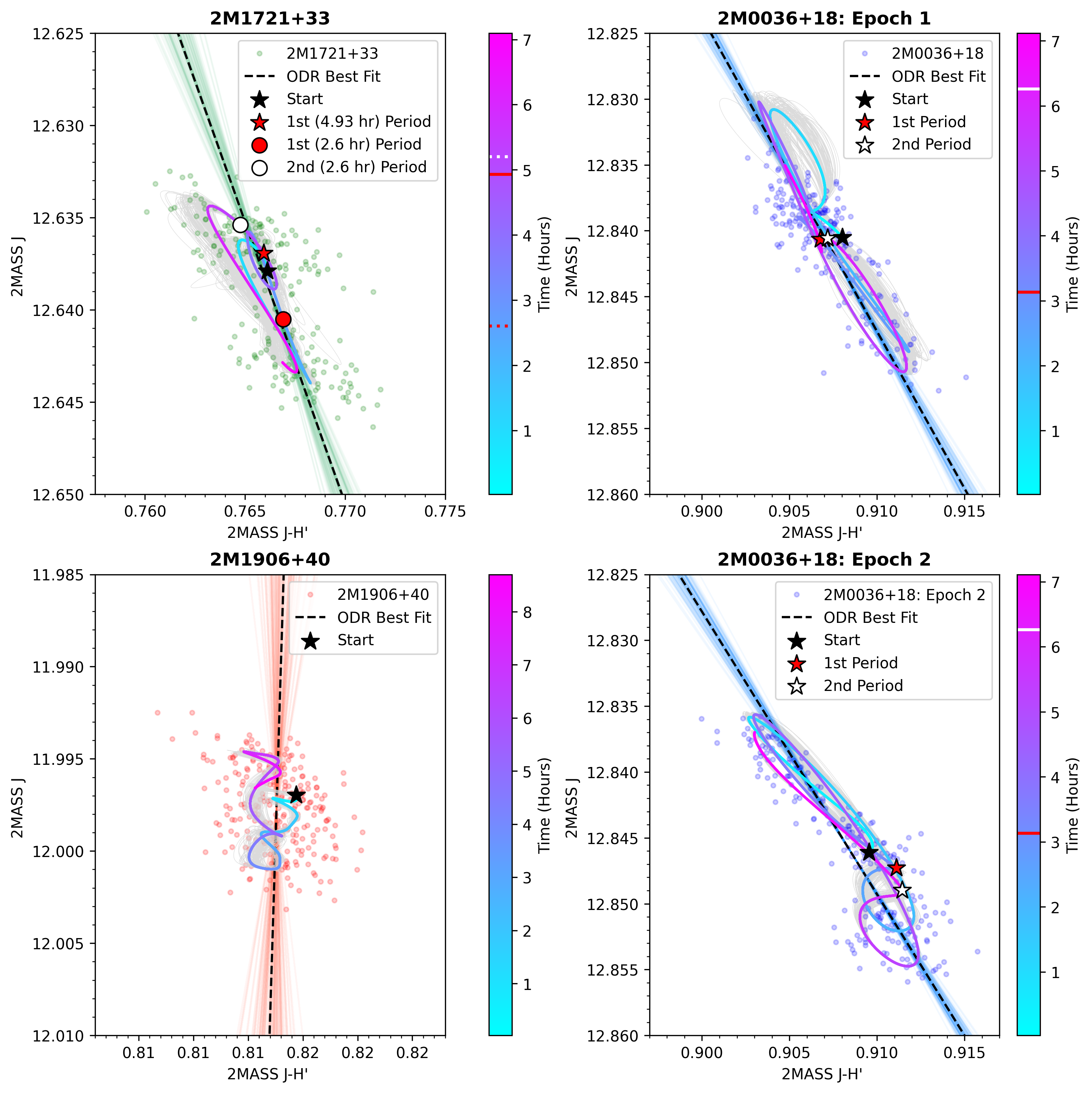}
    \caption{{The colour modulation for each of our objects. Coloured circles represent the HST data shown in the right hand panels of Figure \ref{fig:colour mod}. The dashed and solid lines are the best fit ODR and a selection of 100 posterior ODR fits for each object. The purple-blue solid line represents the colour modulation of the 2MASS J vs 2MASS J-H' {\tt celerité2} light curves. This has the added benefit of interpolating over the gaps left by the HST observations. The colourbar on the right shows how the colour changes as a function of time for each object. The colour modulation of a sample of 100 posterior light curves is also shown for each object in light grey. The location of the start of the observations for each object is highlighted by a black star. A red star is used to highlight the position after one period, and a white star after two periods.
    The start of each period is consistent for each object with more than one rotation, further supporting our chosen periods.
    The ODR linear fits for each object are consistent with the path traced out by the light curve modulation. } }
    \label{fig: LC colour mod}
\end{figure*}

\subsection{Light curve fitting}\label{sec: light curve fitting}
We used {\tt celerité2} \citep{celerite1, celerite2} to model each of our light curves. {\tt Celerité2} is an algorithm for fast and scalable Gaussian processes to determine a best-fit model to describe the variability. We used the {\tt RotationTerm} kernel for each of our light curves. This kernel is typically used for modelling stellar rotation \citep[]{celerite_barros_2020, celerite_VHS_2021} and incorporates two stochastically driven damped simple harmonic oscillating terms, one at a given period and one at half that period. 
We used this kernel as it has been demonstrated {successfully} on brown dwarf light curves in the past \citep[]{celerite_VHS_2021, Mccarthy_2024A, McCarthy2025_JWST_SIMP}. {It is a data driven model that does not take into account the underlying physics of the variability mechanism, simply providing a fit to the data.}
There are five free parameters in this Gaussian process model:  
\begin{itemize}
    \item ${\sigma}$ : The standard deviation of the process.
    \item Period: The primary period of variability.
    \item Q0: The quality factor for the secondary oscillation.
    \item dQ: The difference between the quality factors of the first and second modes
    \item f:  The fractional amplitude of the second mode compared to the primary. 
\end{itemize}

We then used the Markov Chain Monte Carlo (MCMC) package {\tt emcee} \citep{emcee} to find the best-fit model for each of the G141 white light curves{, as well as the 2MASS J and 2MASS H' light curves}.
We used 35 walkers, with a burn in of 2000 steps followed by 10000 steps.
We verified that the chains had converged by visually inspecting the parameter traces.
The parameter uncertainties were taken from the 16th and 84th percentiles of the posterior distributions.

We selected the maximum likelihood model as the best fit light curve for each target.
We show the best-fit models and a sample of 100 posterior fits for each target in the left panel of Figure \ref{fig:LCs+period}.
We calculated the amplitude of each light curve in the same manner as \citet{Wilson_LC_Amp_2014,Radigan_LCs_2014,Biller_2024}. We define the maximum deviation for each light curve as the percent variation between the highest peak and deepest trough over the course of each HST observation. 
The amplitudes are shown in Table \ref{tab: periods}.

We also compute periodograms from our light curves using the astropy Lomb-Scargle function, \citep{astropy:2013,astropy:2018,astropy:2022}.
In the right panel of Figure \ref{fig:LCs+period} we show the periodogram of each data set with the period from our best fit models in black, and the literature period in purple.
The rotation period of 2M0036+18 from our model agrees with the rotation period estimated by  \cite{croll2016long}.

However, we find that for 2M1721+33, our model favours a period of $4.9^{+0.4}_{-0.2}$ hours, in disagreement with the \cite{metchev2015weather} value of 2.6 $\pm$ 0.1 hours. 
The \cite{metchev2015weather} results were based on Spitzer $3.6~\upmu$m and $4.5~\upmu$m photometric light curves of 2M1721+33 over 20 hours of observations. Their light curve model was a truncated Fourier series. 
They varied the number of terms, \textit{n}, in their Fourier series until the residuals were consistent with random noise.
To investigate the period of 2M1721+33, we used \texttt{celerité2} to fit a model to the \cite{metchev2015weather} Spitzer light curve, as seen in the top of Figure \ref{fig:Spitzer LC}, and only applied it to the 3.6$~\upmu$m light curve as the 4.5$~\upmu$m light curve was not reported to be variable.
From this model, we obtain a period of $5.2^{+0.7}_{-1.6}$ hours {for the Spitzer 3.6$~\upmu$m light curve}. 
This period is consistent within uncertainties with the period that we obtained from the HST light curves.
The shape of our periodogram for 2M1721+33 in Figure \ref{fig:LCs+period}, replicates the shape of the periodogram from Figure 3 in \citet{metchev2015weather}, with two significant peaks at 2.6 hours and 4.9 hours.
The two peaks have a very similar periodogram power, with the 2.6 hour peak being slightly higher. 
When taking into account both the Spitzer and HST light curve fits,
our models prefer a double peaked light curve with a period of $4.9^{+0.4}_{-0.2}$ hours. This double peaked feature is especially evident when comparing both the Spitzer and HST light curves as shown in Figure \ref{fig:Spitzer LC}. Despite these two light curves being observed over 9 years apart and at different wavelengths, there is a common shape to both light curves. They both exhibit a primary dip, followed by a secondary dip in each period. The primary and secondary dips in the light curve are seen across both light curves.
This double peaked feature can be explained by the presence of two different atmospheric structures, one in either hemisphere \citep{Vos2018}. {We show how the 2M1721+33 light curve looks when phase folded according the \cite{metchev2015weather} period and the period from this work}.

We found that our period for 2M1906+40, $19.4^{+46.5}_{-10.5}$ hours{,} is poorly constrained. 
Our data does not have a sufficient time baseline to constrain the period and therefore we assumed the period from \citet{gizis2015kepler} for this work.
\citet{Gizis2013} and \citet{gizis2015kepler} both measured 2M1906+40's period to high precision, {8.88364} hours $\pm ~ 5 \times10^{-5}$ hours, using observations from Kepler. They achieved such high precision on the rotation period due to their observations which spanned $>800$ days of continuous observations, from 28 June 2011 to 08 May 2013, (Programs GO 30101, GO 40004). During this time, the sinusoidal shape of the light curve remained consistent. In comparison, we observe $\sim$98\% of one period in this work.

\section{Rotational modulations on the colour magnitude diagram}\label{sec:colour modulation}
The colour magnitude diagram provides a convenient way to compare the broad observational properties of brown dwarfs at a glance. We can also investigate how the colours of brown dwarfs with rotational modulations vary by plotting their colour modulations on this diagram. 
By fitting a line to their J vs J-H' colour, we can measure how the colour of a brown dwarf changes as it rotates.
We calculated the J and H' band magnitudes at each timestamp for each object following the procedure presented in \cite{Lew_colour}.
The full H band range extends past the range of the HST WFC3 instrument ($\sim 1.48-1.82 \upmu$m). Therefore, we define H' as a new band, which only extends to 1.67~$\upmu$m{, the H-band region that is observable by WFC3}. This is consistent with the method in \cite{Lew_colour}. We show the J band magnitudes against the J-H' magnitudes to display the colour modulation in Figure \ref{fig:colour mod}.

Previously, \cite{Lew_colour} measured the J vs J-H' slopes for a sample of 12 brown dwarfs from spectral types L5-T8. These 12 brown dwarfs can be seen, plotted in black, along with our three brown dwarfs in the left panel of Figure \ref{fig:colour mod}.
Following \cite{Lew_colour}, we used orthogonal distance least squares regression (ODR), using the {\tt scipy.odr} package to estimate the slope of the variability in colour-magnitude space as a function of rotation phase. ODR uses the orthogonal distance to define the best fit slope of the modulations across the rotation period, accounting for uncertainties in both dimensions. 
In the right panel of Figure \ref{fig:colour mod} we show the J vs J-H' colour modulation for each of the brown dwarfs from our observations, along with their ODR fits. We fit for $\theta$, the angle our line makes with the horizontal. A positive slope ($\theta < 90^{\circ}$) suggests that the object becomes brighter and redder, while a negative slope ($\theta > 90^{\circ}$) suggests that the object becomes brighter and bluer as it rotates.
While some of the \cite{Lew_colour} {objects} have large J-band variations, i.e. 2M2139, all of the ODR slopes, are very steep. $\theta$ varies from $94^{\circ} < \theta < 103^{\circ} $ among L-dwarfs, and $76^{\circ} < \theta < 97^{\circ} $ among T dwarfs. This indicates that despite large variability amplitudes, their colour does not change significantly with phase. 

Adding our three early L dwarfs to this sample allows us to probe a complementary parameter space of hotter effective temperatures. The slopes of the ODR fits to the three brown dwarfs follow an interesting trend: both 2M1721+33 and 2M0036+18 ({epoch} 1 and {epoch} 2) exhibit considerably shallower slopes than the other brown dwarfs in this sample {($\theta = 108.1^\circ\,^{+1.7^{\circ}}_{-1.8^{\circ}}$,$\theta = 112.7^\circ\,^{+1.0^{\circ}}_{-1.0^{\circ}}$ and $\theta = 115.0^\circ\,^{+1.0^{\circ}}_{-1.0^{\circ}}$ }respectively). The $\theta$ of both observations of 2M0036+18 are {similar but are slightly different. This can be attributed to the fact that each epoch observes a slightly different phase of 2M0036+18.} This can be seen in the middle sub panel of Figure \ref{fig:colour mod}. As these two objects rotate they become brighter and bluer in J-H'.
2M1906+40 on the other hand has a drastically different slope compared to the other two objects in this sample, with $\theta = 88.5^\circ\,{^{+5.7^{\circ}}_{-0.2^{\circ}}}$. This results in a near vertical slope, indicative of grey variability modulations. It is also the steepest slope among all the L-dwarfs between this and \cite{Lew_colour}'s sample. {We note that we use these slopes as tools to help illustrate the trend in the colour modulation of each object.} 

{In order to further justify our fits to the colour modulation, we use the approach described in \cite{Mccarthy_2024A}. We plotted the colour modulation of the 2MASS J and the 2MASS J-H' {\tt celerité2} light curves. This is particularly useful for 2M1721+33 and 2M1906+40, where the trend in colour modulation is not immediately obvious. As illustrated in Figure \ref{fig: LC colour mod}, the general trend of the light curve modulation matches the ODR fits shown in Figure \ref{fig:colour mod}.
The shape of 2M1721+33's colour modulation is distinctly split into an upper and lower region, as seen in Figure \ref{fig:colour mod}. However, this shape is due to the fact that we do not obtain full phase coverage of 2M1721+33's light curve with HST. As illustrated in Figure \ref{fig: 1721_phase_fold}, we observe the peaks and troughs of its period, missing out on the middle portion which results in our two regions in colour space. However, when using the interpolated light curve colour modulation in Figure \ref{fig: LC colour mod}, it is evident that our ODR fits follow the same path. Similar colour-modulation shapes have been seen in other HST variability papers \citep[]{Lew_colour,HSTVAR_2025}.}

\citet{Lew_colour} suggest a potential trend in the colour modulation of brown dwarfs across their sample. Early T-dwarfs become brighter with little to no colour change, while L-dwarfs become brighter and bluer. Our sample allows us to investigate this trend at earlier spectral types, and we find that the trend continues at these warmer temperatures. Both 2M1721+33 and 2M0036+18 become brighter and bluer than any other brown dwarf in the {\cite{Lew_colour}} sample. Whereas 2M1906+40 has a near vertical, positive slope that is steeper than any other L-dwarf, {with grey modulations}. The behaviour of our three objects in Figure \ref{fig:colour mod}, suggests that brown dwarfs continue to become brighter and bluer as they rotate, except for 2M1906+40 which has grey modulations.

\begin{figure*}[t]
    \centering
    \includegraphics[width= \textwidth]{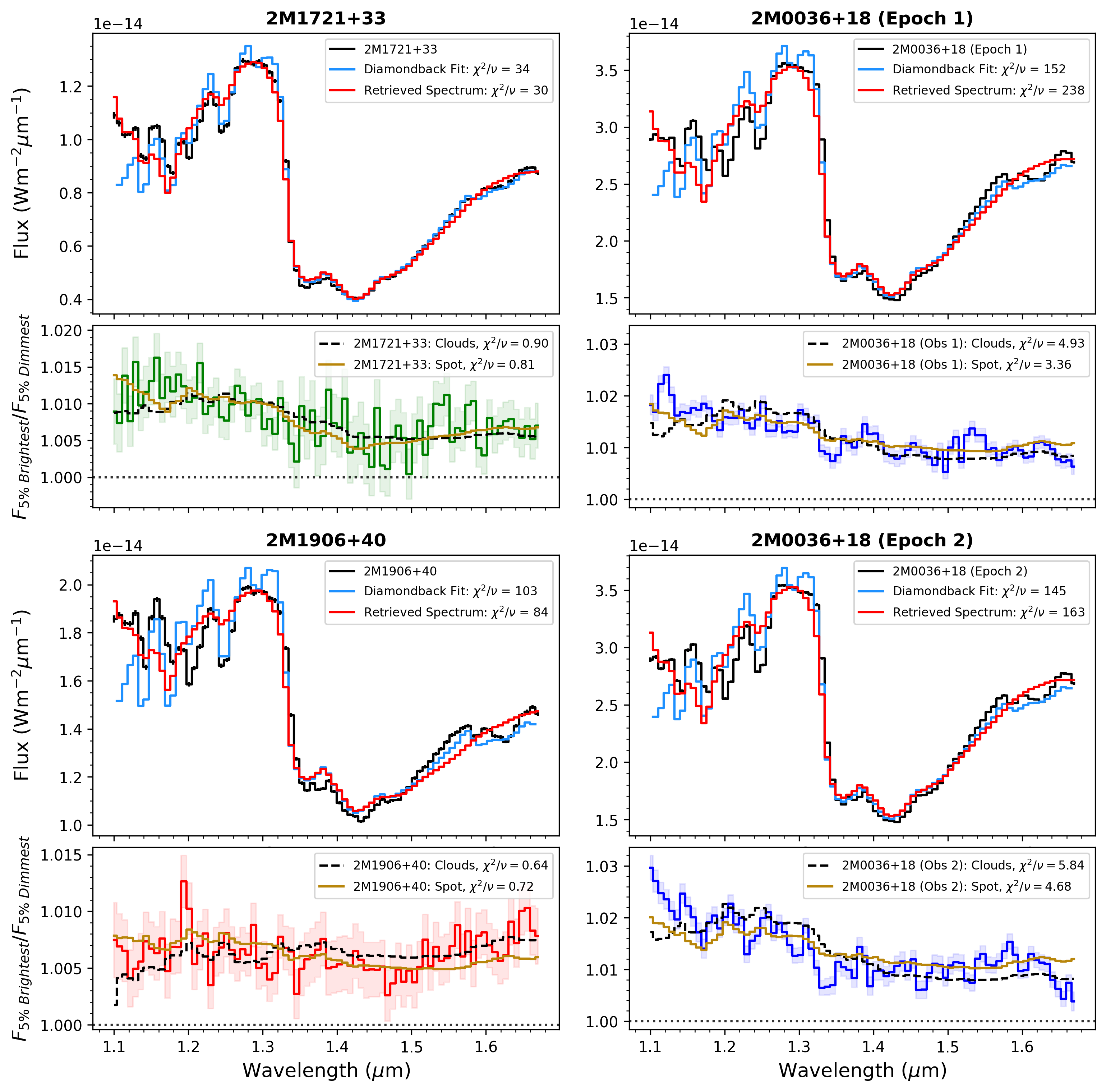}

    \caption{The HST spectrum (black line), Sonora Diamondback best fit spectrum (blue line) and petitRADTRANS retrieval spectrum (red line) is in the upper panel for each plot. The bottom panels shows the spectral variability amplitude for each object (green, blue or red lines). The cloud and spot driven spectral variability models are plotted in black and gold respectively. For both 2M1721+33 and 2M1906+40, the cloud-driven and magnetic spot-driven models both provide adequate fits to the data. For both observations of 2M0036+18, there is a slightly more favourable $\chi^{2}/ \nu$ value for the spot model over the cloud model. }
    \label{fig:cloud_mode+spot_models}
\end{figure*}

\section{Establishing and modelling the spectral variability amplitude}\label{sec:spectral variability amplitude}
The spectral variability amplitudes (wavelength dependence of the variability) allow us to discern the main drivers of variability on brown dwarfs. Within the HST wavelengths, the shape has been found to differ across spectral types, with L dwarfs typically exhibiting amplitudes that decrease smoothly as a function of wavelength \citep[]{yang2015, Manjavacas2018,Lew2020CloudDwarf}. However, in L/T transition objects, the water band is typically less variable than the surrounding continuum \citep{yang2015, Zhou2018}. 
The reason for this change is that in the L-dwarf regime, there are spatially varying high-altitude clouds that drive the variability \citep{yang2015}. In contrast, along the L/T transition the condensate clouds are deeper in the atmosphere, which results in a reduction of the variability amplitude within the water opacity band \citep[]{yang2015, Vos2023PatchyAnalogs}. 

The spectral variability amplitude of each observation was obtained by calculating the ratio of the 5\% brightest spectra to the 5\% dimmest spectra in each observation. The brightest and dimmest 5\% of spectra were chosen because they are representative of these regions while also being a typical amount used in the literature ($\sim 5-10\%$) \citep{Zhou2018, Lew_colour}. We chose this percentage as it boosts the signal to noise as opposed to simply taking the single brightest and dimmest spectra, while also still probing a small percentage of the phase.
We used the {\tt spectres} package \citep{carnall2017spectresfastspectralresampling} to resample the wavelength points onto a common grid and interpolate the flux across these points before calculating the brightest/dimmest ratio. The error on this ratio was calculated using MCMC bootstrapping.

The spectral variability amplitudes of our objects are shown in Figure \ref{fig:cloud_mode+spot_models}. All three targets are variable at all wavelengths. It is clear that they are similar to other HST observations of brown dwarfs in the L-dwarf regime. They do not exhibit lower variability amplitudes in the water band, as is typical of L/T transition objects \citep{Apai_2013, yang2015}.
The variability for 2M1721+33 and both observations of 2M0036+18 exhibit a slight slope, where the amplitude is highest at bluer wavelengths and decreases at longer wavelengths. In contrast, the variability of 2M1906+40 shows little wavelength dependence. 2M0036+18 exhibits the highest variability amplitude (in both observations), followed by 2M1721+33 and then 2M1906+40.

\subsection{Modelling the spectral variability amplitude}

Clouds, aurorae and magnetic spots are three possible drivers of variability for early L-dwarfs. Each of these processes has a different effect on the spectral variability amplitude.
To model the spectral variability amplitude, we use a combination of radiative-convective equilibrium forward models, together with a more flexible atmospheric retrieval approach.

Multiple modelling and retrieval studies have shown that having a wide wavelength range is critical for determining the thermal profile \citep{Burningham2021,Vos2023PatchyAnalogs}.
As our observations cover a narrow wavelength range, we use the pressure temperature profile from the best fit Sonora Diamondback grid models \citep{Morley2024TheObjects}. In order to obtain the pressure temperature profiles,
we used the Sonora Diamondback grid models to identify the best fitting spectrum to the median spectrum of each object.
Sonora Diamondback is a one-dimensional fully self-consistent model with five varying parameters: temperature, gravity, a cloud sedimentation parameter $f_{sed}$, metallicity and the C/O ratio relative to solar values.
We linearly interpolated these models over the grid to find the best fit using the {\tt{species}} and  {\tt{pymultinest}} packages \citep{species,pymultinest}. 
The optimal fits, shown in Figure \ref{fig:cloud_mode+spot_models}, have reduced $\chi^{2}$ values of 34, 152, 145 and 103 for 2M1721+33, 2M0036+18 ({epoch} 1), 2M0036+18 ({epoch} 2) and 2M1906+40 respectively. From the interpolated best fit, we took the closest model from the grid and used its pressure temperature profile for the atmospheric retrievals. The parameters for these models are found in Table \ref{tab: Retrieval Priors}, and the fits themselves in Figure \ref{fig:cloud_mode+spot_models}.

Clouds are expected to form in early L-dwarfs \citep{Burrows2006}. We compared the condensation curve of forsterite (Mg$_{2}$SiO$_{4}$) to the pressure temperature profile of each of our objects (taken from the Sonora Diamondback models) and found that it is expected to form in the photosphere of each of our objects (1-0.1 bar). The justification for clouds is further strengthened since both 2M1721+33 and 2M0036+18 show prominent silicate features at 10$\upmu$m in archival Spitzer IRS data \citep{Cushing2006,Suarez2022}.

\begin{figure}[t]
    \centering
    \includegraphics[width=\columnwidth]{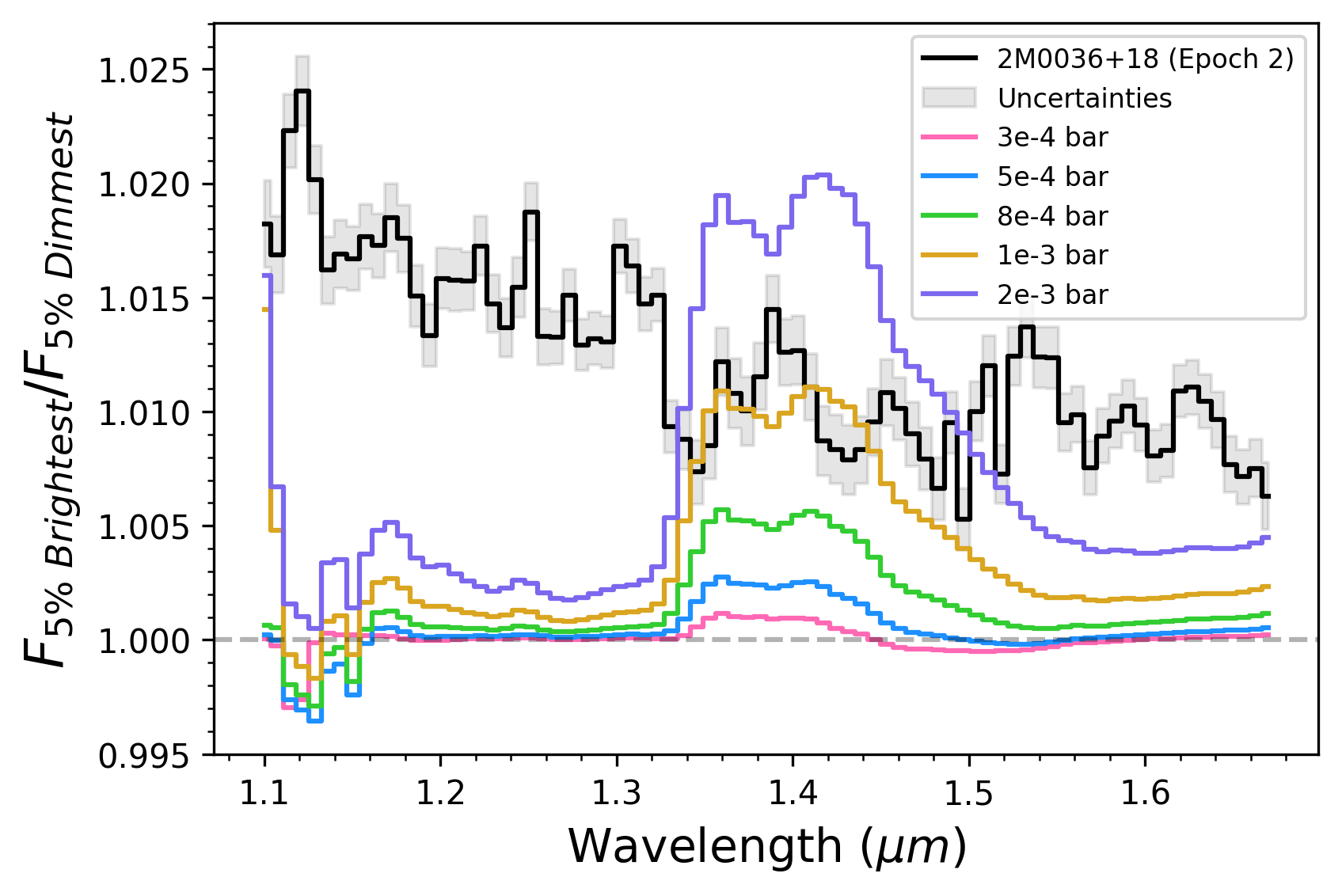}
    \caption{Spectral variability amplitude of 2M0036+18 ({epoch} 2) (black line with uncertainties in grey) with the aurora modelled by the temperature inversion method. This model is for a 350K temperature inversion at different peak pressures, with a width of 1 pressure scale height (10-0.1 mbar) and a $\Delta$A = 0.1. The models are shown by the coloured lines. As the peak pressure of the inversion is placed at higher pressures, lower in the atmosphere, the predicted variability amplitude in the water band increases. However, there is very little variability outside the water band, no matter the temperature inversion, or peak pressure value. }
    \label{fig: Aurora Model}
\end{figure}

\begin{table*}
\centering
\caption{Uniform prior ranges and results from the petitRADTRANS retrievals and the parameters for the best fit Sonora Diamondback spectra.}

\renewcommand{\arraystretch}{1.2}

{\begin{tabular}{llllll}
\toprule
\textbf{Parameter} & \textbf{Prior} & \textbf{2M1721+33} & \textbf{2M0036+18 ({epoch} 1)} & \textbf{2M0036+18 ({epoch} 2)} &  \textbf{2M1906+40}\\
\midrule

\textbf{petitRADTRANS }   
  
\\
\midrule

\textbf{Log (g)} & $\mathcal{U}(3.0, 5.5)$  &  4.24$\pm$0.04   &  4.27$\pm0.01 $ & 4.27$\pm0.01 $ & 4.96$\pm$0.06 \\

\textbf{Radius ($R_{Jup}$)} &$\mathcal{U}(0.6, 1.4)$  &  0.84$\pm$0.01  & 0.81$\pm$0.01  & 0.81$\pm$0.01 & 0.85$\pm$0.01\\

\textbf{Cloud Fraction}  ($f_{cloud}$) & $\mathcal{U}(0, 1)$ &   0.73$\pm$0.01 &   0.83$\pm$0.01 &  0.83$\pm$0.01  &   0.82$\pm$0.01 \\

\textbf{k$_{0}$} (350nm opacity) & $\mathcal{U}(-5, 20)$  &  2.42$\pm0.75$  & 2.01 $\pm$0.04 & 2.02 $\pm$0.04 &   5.62$\pm$0.68\\

$\boldsymbol{\gamma}$ (Power Law exponent) & $\mathcal{U}(-20, 20)$  &   -4.23$\pm$0.20  &  -4.02$\pm$0.02  &  -4.04$\pm$0.02 &  -4.01$\pm$0.02 \\

\textbf{Log$_{10}$ CO} & $\mathcal{U}(-7, 0)$  &   -2.54$\pm$1.97 &  -2.60$\pm$0.03 & -2.61$\pm$0.03  &   -1.99$\pm$0.07 \\

\textbf{Log$_{10}$} $\mathbf{H}_{\mathbf{2}}{\mathbf{O}}$ & $\mathcal{U}(-7, 0)$ &   -3.09$\pm$0.02  & -3.22$\pm$0.01 &  -3.22$\pm$0.01  &   -2.79$\pm$0.05  \\

\textbf{Log$_{10}$ K} & $\mathcal{U}(-7, 0)$  &  -6.85$\pm$ 0.04  & -7.00$\pm$0.01 & -6.99$\pm$0.01 &   -7.00$\pm$0.01 \\

\textbf{Log$_{10}$ Na} & $\mathcal{U}(-7, 0)$ &  -6.98$\pm$ 0.04  & -7.00$\pm$0.01 & -7.00$\pm$0.01 &   -6.98$\pm$0.03 \\

\midrule

\textbf{Sonora Diamondback}    

\\
\midrule

\textbf{$T_{eff}$ (K) } & - & 1700 & 1600 &  1600 & 2000  
\\

\textbf{Log (g)} & - & 5 & 5&5  &  5.5
\\

\textbf{Metallicity} & - & +0.5 &+0.5 & +0.5 1.1&  +0.5
\\

\textbf{C/O} & - & 1.0 & 1.0& 1.0 &  1.0
\\

\bottomrule
\end{tabular}}
\tablefoot{The abundances of the species are in units of log mass mixing ratio.}
\vspace{2mm} 
\label{tab: Retrieval Priors}
\end{table*}

We used atmospheric retrievals to model the median spectrum of each object as they provide greater flexibility for the parameters to reproduce the spectral variability amplitudes. Self-consistent grid models do not have the resolution in parameter space to capture the $\sim 1\%$ variability that we are trying to fit.
We used the retrieval package \texttt{petitRADTRANS} \citep[]{Molliere2019, Nasedkin2024} to run our analysis.
We fitted the same median spectrum for each object that we used when getting the best fit Sonora Diamondback spectrum.
We fixed the pressure temperature profile to the self-consistent Diamondback profile, as the HST spectra do not have a wide enough wavelength coverage or sufficient spectral resolution to constrain it \citep{Burningham2021,Vos2023PatchyAnalogs}. 
The distance to each object was also fixed to their estimated distances from the literature, (16.18 $\pm ~0.04$ pc, 8.74 $\pm ~0.01$ pc, 16.76 $\pm ~0.03$ pc for 2M1721+33, 2M0036+18 and 2M1906+40 respectively \cite{Gaia2021}). The brown dwarf radius and surface gravity {was} allowed to vary, as well as the atmospheric mass fraction abundances of CO, H$_{2}$O, K and Na \citep{HITEMP_CO_2010, h2O_opacity2018, Molliere2019, Allard_Na_opacity2019}.
The remaining atmosphere was filled with H$_{2}$ and He, forming 76\% and 24\% of the mass respectively, similar to the composition of Jupiter \citep{JUP_H2_2003,JUP_H2_2015}.
We account for Rayleigh scattering in H$_{2}$ and He as well as the collision induced absorption for the H$_{2}-$H$_{2}$ and H$_{2}-$He interactions. We ran the retrievals using nested sampling, using the {\tt pyMultiNest} package \citep[]{Feroz2009,Feroz_2013, Buchner2014}.

Three parameters were used to model clouds in the atmosphere. A cloud fraction ($f_{cloud}$) parameter which parametrises to what extent the brown dwarf's atmosphere is covered by clouds. We also had two power law cloud parameters. The first power law cloud parameter is the opacity of the cloud at 350 nm, k$_{0}$, and the second is the power law cloud exponent, $\gamma$. 
This parametrisation was chosen as it approximates the effect of clouds by introducing a power law opacity slope to the spectrum. $\gamma$ typically ranges from -5 to 0, with the approximated scattering increasing with increasing $\gamma$, as the opacity slopes also increase. The value $\gamma$ characterises how \say{non-grey} the clouds are \citep{Burningham2017}, with decreasing values of $\gamma$ resulting in increased non-greyness of the clouds and with $\gamma=0$ mimicking a grey cloud deck.

We tested clear, patchy, and fully cloudy atmospheres, and found that using a patchy power law cloud fit the spectral variability amplitudes the best.
A patchy cloud model{, where the clouds were grey,} failed to reproduce the spectral variability amplitude as accurately as the patchy power law model. The $\chi^{2}/\nu$ value for these models was at least double that of the best fit patchy power law cloud model that we used. We therefore favoured the patchy power law model.

We used 400 live points and had an evidence tolerance of 0.5 in our retrievals. We calculated our models at a spectral resolving power of 180 before convolving them to the spectral {resolving power} of the G141 spectra, $R\sim 130$. They were then binned to the wavelength grid of the data to calculate the log likelihood.
We set broad priors on our retrieved parameters, as shown in Table \ref{tab: Retrieval Priors}. We also retrieved mass fraction abundances for each of the species in our atmosphere. The best fit retrieved spectra are shown in Figure \ref{fig:cloud_mode+spot_models}.

We chose to model the spectral variability amplitude of each object by creating two separate regions in the atmosphere. For each different variability mechanism, we created a different atmospheric region. We modelled cloud driven variability by changing the properties of the clouds in each region. We modelled aurorally driven variability by inserting a temperature inversion into the upper atmosphere of one of our regions. This {temperature inversion} follows findings from \citep{Faherty2024MethaneDwarf, nasedkin2025} where aurorae can heat the upper atmosphere. We modelled magnetic spot driven variability by reducing the temperature of the pressure temperature profile, in order to mimic the effects of a magnetic spot \citep{Rockenfeller2006, Barnes2015, Rackham2018}.
Other than each specific change for the different variability mechanisms, the models were identical to the best fit retrieval.
By altering our best fit retrieval with each of these parameters (changing the cloud properties, inserting a temperature inversion, and reducing the temperature of the pressure temperature profile), we can  model the spectral variability amplitude.

We used the following equation, adapted from \cite{Lew2020CloudDwarf}, to model the spectral variability amplitude. 

\begin{equation}  
    V = 1 + \Delta A  \left( \frac{F_{brighter}-F_{dimmer}}{F_{min}} \right )
    \label{eqn: wvar modelling}
\end{equation}

Where $\Delta A$ is the coverage fraction of the secondary region (i.e. the region with differing cloud properties, an auroral temperature inversion or a magnetic spot). $F_{brighter} - F_{dimmer}$ is the difference between the flux of the brighter and dimmer spectra between the baseline retrieved spectrum and the secondary spectrum. $F_{min}$ is the observed spectrum of the mean 5\% dimmest spectra of the object. \textit{V} is the variability amplitude as a function of wavelength.

\subsection{Cloud modelling}
To model cloud driven variability, we assume that the variability is caused by two regions in the atmosphere with different cloud properties. The variability was modelled by changing the values of both k$_{0}$, the opacity at 350 nm, and $\gamma$, the power law opacity exponent. 
We fit the spectral variability amplitude using Equation \ref{eqn: wvar modelling}.
We ran an MCMC using \texttt{emcee} to optimise our model and to constrain our uncertainties \citep{emcee}, and the median value of our constrained parameters for each model are shown in Table \ref{tab: cloud_models}.
We used 20 walkers and 500 steps for each observation until the model converged. As the two power law cloud parameters are degenerate, there is a spread in the possible values for a good fit, as shown in Appendix \ref{app: Appendix_Cloud_corners}, Figures \ref{fig: 1721_corner}, \ref{fig: 0036_obs1_corner}, \ref{fig: 0036_obs2_corner}, and \ref{fig: 1906_corner}.

We show our model fits for cloud driven variability in Figure \ref{fig:cloud_mode+spot_models}.
Our model for cloud driven variability fits the spectral variability amplitude for 2M1721+33 and 2M1906+40, with $\chi^{2}/ \nu$ < 1 for both cases, and a slightly poorer fit for 2M0036+18. The higher $\chi^{2}/ \nu$ for 2M0036+18 is driven by its smaller uncertainties.
Our model follows the general shape of the spectral variability amplitude for all three of our objects.
These models indicate that the observed spectral variability may be produced by two regions in the atmosphere that host clouds with differing properties.

\subsection{Auroral modelling}

We used a similar framework to model auroral variability. However, instead of changing the cloud property parameters, we inserted a temperature inversion in the upper atmosphere of the pressure temperature profile to create our secondary auroral region. 
Temperature inversions are common in our solar system, and in the case of Jupiter, the bulk of this upper atmosphere heating is driven by the redistribution of heat from hot polar auroral regions \citep{odonoghue2021}.
Evidence of temperature inversions in brown dwarf atmospheres has recently been reported by \citet{Faherty2024MethaneDwarf}, where auroral emission of methane was reproduced by inserting a temperature inversion into the pressure temperature profile. 
However, unlike \citet{Faherty2024MethaneDwarf}, we assume that the auroral feature only exists in one region of our atmosphere.
Following examples by \cite{Morley2014SpectralDwarfs} and \citet{nasedkin2025}, we insert a Gaussian temperature perturbation into the pressure temperature profile. The perturbation has three parameters: the temperature of the inversion, the pressure of the peak of the inversion and the width of the temperature inversion. We insert a temperature inversion of 350~K with a peak pressure of 1 mbar and a width of one pressure scale height (10-0.1 mbar), similar with retrieved values from \citet{Faherty2024MethaneDwarf, nasedkin2025}. While our selection of a 350~K inversion is somewhat arbitrary, we do not have the wavelength coverage nor resolution to perform retrievals as in \citet{Faherty2024MethaneDwarf, nasedkin2025} to obtain a more accurate inversion value. However, the effects of the temperature of the inversion are simply more prominent with a higher temperature. We tested the effect of lower temperature inversion (100 K, 200 K) on the model but it resulted in the same shape of the model, but at a lower amplitude.
350~K helps to visualise these effects while also being similar to those in the literature. 

As before, we fit the spectral variability amplitude using Equation \ref{eqn: wvar modelling}.
We modelled two regions of the atmosphere. The primary region is the fiducial state of the atmosphere as measured by the atmospheric retrieval, while the second region includes the auroral temperature perturbation. 
In Figure \ref{fig: Aurora Model} we show the variability amplitude predicted by a 350K inversion at a range of different pressures in the atmosphere, with $\Delta A$ = 0.1.
The majority of the variability is found within the water band at $\sim 1.3-1.45 ~ \upmu$m, as shown in Figure \ref{fig: Aurora Model}. When the temperature inversion is located at lower pressures, the amplitude of the variability is weaker. However, as the location of the temperature inversion moves deeper in the atmosphere, the amplitude of variability within the water band increases.
This model clearly does not reproduce the spectral variability of 2M0036+18 across the full HST wavelength range.

\begin{table*}[ht]
\centering
\caption{Best fit parameters for the cloud and magnetic spot models for each brown dwarf.}

\renewcommand{\arraystretch}{1.2}

{\begin{tabular}{lcccccc|cccc}
\toprule
\makecell{\textbf{Object}} & \thead{\textbf{Retrieved $\gamma$}} & \thead{\textbf{Model $\gamma$}} &
\thead{\textbf{Retrieved  k$_{0}$}} & \thead{\textbf{Model k$_{0}$}} & 
\thead{\textbf{$\Delta A$}} (\%) & \thead{\textbf{$\chi^{2}/\nu$}} & \thead{\textbf{$\Delta T$ (K) }}  & \thead{\textbf{$\Delta A$}} (\%) & \thead{\textbf{$\chi^{2}/\nu$}} \\
\midrule

 2M1721+33 &   $-4.23^{+0.15}_{-0.20}$  &$-4.11^{+0.07}_{-0.09}$ & $2.42^{+0.75}_{-0.43}$  &  $1.61^{+0.14}_{-0.19}$  &$15.50^{+9.48}_{-6.81}$ & 0.90 & -200 & $2.89^{+0.07}_{-0.08}$ & 0.81 \\
 
 2M0036+18 (Obs 1) & $-4.02^{+0.01}_{-0.02}$ & $-4.00^{+0.02}_{-0.03}$ & $2.01^{+0.04}_{-0.03}$ & $1.46^{+0.04}_{-0.04}$ &$12.95^{+0.20}_{-0.27}$ & 4.93 & -200 &  $2.95^{+0.02}_{-0.03}$ & 3.36 \\

 2M0036+18 (Obs 2) & $-4.04^{+0.02}_{-0.02}$ & $-3.87^{+0.04}_{-0.06}$ & $2.02^{+0.04}_{-0.04}$ & $1.04^{+0.06}_{-0.10}$ &$10.15^{+0.47}_{-2.65}$ & 5.84 & -200 &$3.25^{+0.03}_{-0.03}$ & 4.68\\
  
 2M1906+40 & $-4.01^{+0.01}_{-0.02}$ & $-3.03^{+0.03}_{-0.03}$ &  $5.62^{+0.68}_{-0.49}$ & $5.65^{+0.20}_{-0.22}$ &$2.10^{+0.05}_{-0.06}$ & 0.64 & -300 & $1.46^{+0.04}_{-0.04}$ & 0.72 \\

\bottomrule

\end{tabular}}

\tablefoot{k$_{0}$ is the opacity of the clouds at 350nm and  $\gamma$ is the  power law exponent of the clouds.
$\Delta A$ is the coverage fraction between the bulk atmosphere and the perturbed atmosphere (due to clouds or magnetic spots).}

\vspace{2mm} 
\label{tab: cloud_models}
\end{table*}

\begin{figure}
    \centering
    \includegraphics[width= \columnwidth]{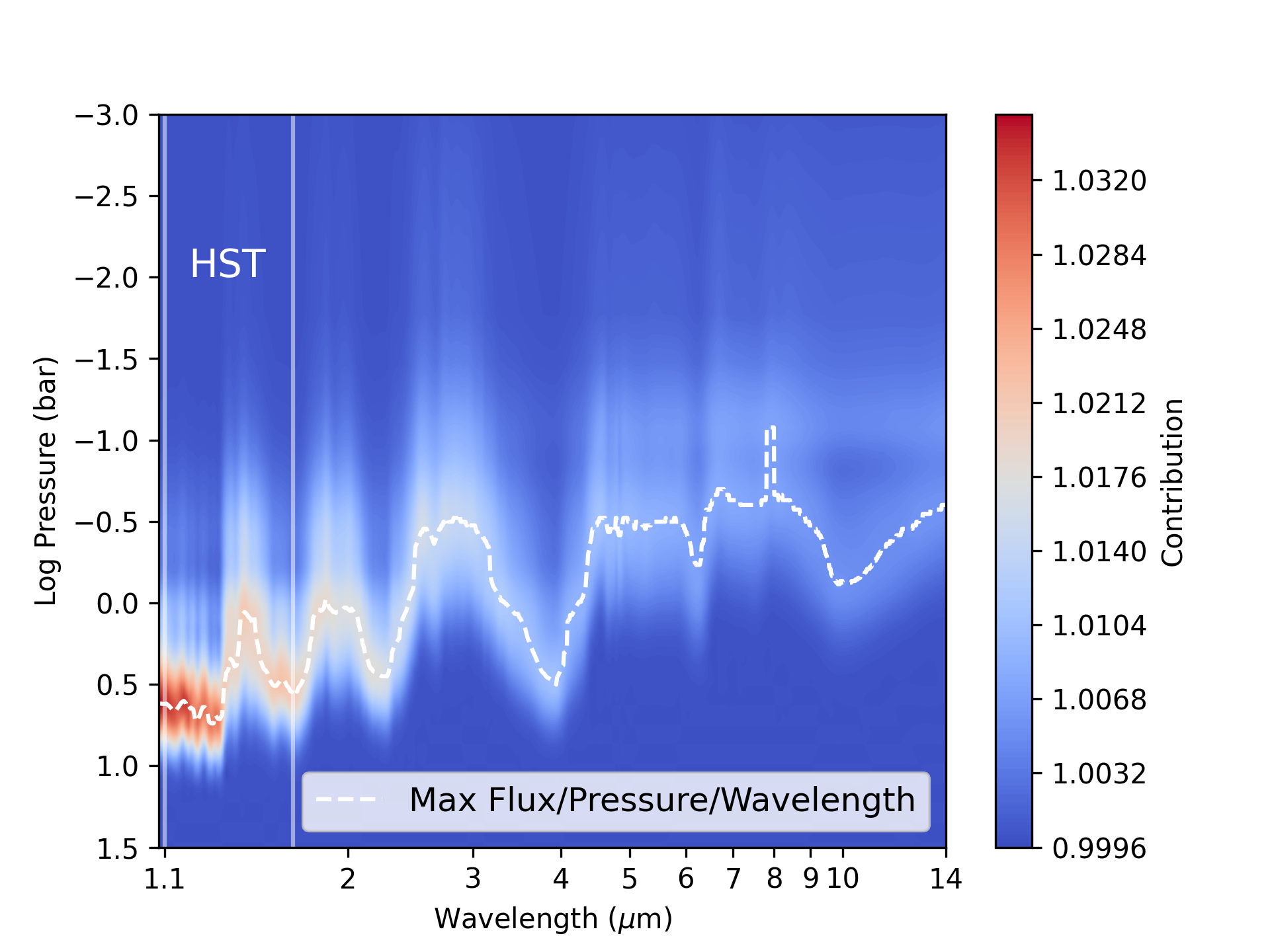}

    \caption{Contribution function for
    a T$_{eff} =$ 1900K, log(g) = 5, $f_{sed}=2$, solar metallicity and solar C/O ratio atmosphere. The pressures probed by the HST wavelengths (1.1-1.67 $\upmu$m), highlighted by the vertical white lines, probe deeper than longer wavelengths that are accessible to JWST. Observations at longer wavelengths that probe the upper atmosphere may be more sensitive to auroral signatures.
    }
    \label{fig: cont}
\end{figure}

\subsection{Magnetic spot modelling}

Magnetic spots are regions on stellar/substellar objects where the magnetic field suppresses the emission of flux from the surface, analogous to sunspots on the Sun. They are common features on M dwarfs and are postulated to be common on L-dwarfs too \citep{Ldwarf_mag_spot2015}. By changing the emitting flux, such spots are hypothesised to be potential drivers of variability \citep{croll2016long}.

Magnetic spot driven variability has been modelled on M dwarfs in the literature by using cooler and hotter regions in the models. These typically consist of a bulk photosphere region, with a higher T$_{eff}$, and a magnetic spot region with a cooler T$_{eff}$ \citep[]{Afram2015,Rackham2018}. 
To model {the spectral} variability {driven by} magnetic spots, we altered the pressure temperature profile of our best-fit forward model. {Our first region represents the bulk photospheres and is the same as in the previous models -- our unperturbed retrieved spectrum for each object.}
Our second region {mimicks} the {effects of} magnetic spot{s with} a Sonora Diamondback PT profile that is 100-300 K cooler for each object. This is in line with the {$\Delta$ T$_{eff}$ values between} magnetic spots {and the bulk T$_{eff}$} on mid to late M dwarfs \citep{Rockenfeller2006, Barnes2015}. \citet{Rockenfeller2006} report variability as a result of a magnetically induced cool spot on the M9 dwarf 2M1707+64, with a T$_{eff}$ 100K cooler than the T$_{eff}$ of the bulk atmosphere.

All other parameters were kept constant. This created two regions of the atmosphere, and using our model from Equation \ref{eqn: wvar modelling}, we can reproduce magnetic spot induced variability. 
{This method of modelling the effects of magnetic spot variability is common, particularly for M-dwarfs hosting exoplanet systems. For example, \citet{Afram2015} modelled the effects of starspots on M dwarfs and found that the ratio between the photosphere and the spots temperatures, $\frac{T_{spot}}{T_{phot}} \approx 0.86$. Using this temperature ratio relation, \cite{Rackham2018} found that the typical covering fraction of spots on M9 dwarfs to range from $1\% < f_{spot} < 24\%$. We note that while this is a simple model to replicate the effects of magnetic spot induced variability, we are limited by the nature of our observations, which are low resolution spectra covering a wavelength range of 0.57 $\upmu$m. In order to produce a more sophisticated model (i.e. one that could Doppler image the spots and highlight the latitude and longitude of the spots) would require high resolution spectra \cite{Roettenbacher_2017}.}

The best-fit parameters for each observation are found on the right hand side of Table \ref{tab: cloud_models}. We ran an MCMC to obtain the best-fit $\Delta A$ across each of the 100-300 K cooler magnetic spots. This used the same set up as for the cloud models, (20 walkers, 500 steps). Using this MCMC we then found the optimal $\Delta T$ and subsequently $\Delta A$ values for each observation. We chose the $\Delta T$ value with the lowest $\chi^{2}/ \nu$ value. The posterior distribution for the $\Delta A$ value for the best fit $\Delta T$ for each object is shown in Figure \ref{fig: spot_mcmc}.

Figure \ref{fig:cloud_mode+spot_models} shows our magnetic spot driven variability model.
This paper is the first to investigate the presence of magnetic spot induced variability on brown dwarfs using HST.
As with our cloud driven variability, the magnetic spot variability models fit our data very well. For each observation, using a spot with $\Delta T_{eff}=-200~K$ provided the best fitting model, except for 2M1906+40, where $\Delta T_{eff}=-300~K$ provided the best fit, with the{se $\Delta T_{eff}$ values providing the} lowest $\chi^{2}/\nu$ value for each object. 
Again, these models fit 2M1721+33 and 2M1906+40 slightly better than 2M0036+18, and this can be attributed to 2M0036+18's smaller uncertainties.
{These results are consistent with the M dwarf studies of \citep{Afram2015,Rackham2018}. The ratio between the temperature of the photosphere and the spots are similar to the 0.86 value of \cite{Afram2015}, ranging from 0.85-0.88 for each object and the coverage fractions (analogous to the $\Delta A$ values presented in this work), also all lie between $1\% < f_{spot} < 24\%$.  These results indicate that our model results are at least plausible as they are consistent with the literature. }

\section{Discussion}\label{sec: discussion}
\subsection{Comparison between the cloud, auroral and magnetic spot models}

We found that our cloud and magnetic spot  models fit the spectral variability amplitude the best for each of our objects. They have similar $\chi^{2}/\nu$ values for each object, so from our observations we cannot rule out one model over another. Future observations at longer wavelengths are required to disentangle these two processes. Both 2M1721+33 and 2M0036+18 exhibit a 10 $\upmu$m silicate feature in their spectra \citep{Cushing2006,Suarez2022}, which indicates that silicate clouds are present in the{ir} atmosphere{s}. Variability at this 10 $\upmu$m silicate feature has been a proposed indicator of cloud driven variability \citep{Biller_2024, McCarthy2025_JWST_SIMP, Chen2025}. Probing the variability at these wavelengths could determine whether the variability is primarily driven by clouds or magnetic spots.
If the variability amplitude continues to decrease at longer wavelengths, that may also be indicative of magnetic spots, as seen in starspot variability on M dwarfs \citep{Rockenfeller2006}.

As shown in Figure \ref{fig: Aurora Model}, our auroral model did not fit the shape of the observed spectral variability. 
In Figure \ref{fig: cont} we show the contribution function for a T$_{eff} = $ 1900K, Log(g) = 5, $F_{sed}=2$, solar metallicity and solar C/O ratio atmosphere, which closely match the fundamental parameters derived for 2M0036+18 by \citet{filipazzo2015}.
We use the \cite{filipazzo2015} parameters for the contribution function instead of our own from the HST data, as their parameters were derived from analysis of the spectral energy distribution over a broader wavelength range, including SPEX-SXD \citep{Spex_0036_2009} and Spitzer IRS spectra \citep{Cushing2006}.
The contribution shows that the majority of the HST flux (wavelengths of 1.1-1.67 $\upmu$m) originates between 1-10 bars. In contrast, at longer wavelengths, the flux originates at shallower pressures, even probing the millibar pressures where temperature inversions have been detected, which may be more sensitive to auroral chemistry \citep{Faherty2024MethaneDwarf}. This is particularly noticeable from $\sim$4.5 - 5.5 $\upmu$m and from $\sim$7.5 - 8.5 $\upmu$m. Observations at these wavelengths will be more sensitive to auroral effects as they probe depths that are associated with a temperature inversion \citep{Faherty2024MethaneDwarf}. We note that the wavelengths that reproduce the variability in our auroral model in Figure \ref{fig: Aurora Model}, $\sim 1.32 - 1.50~ \upmu$m, is the only {region} of the contribution function that probes higher in the atmosphere.

When probing the upper atmosphere at longer wavelengths, we are more sensitive to the potential effects of auroral chemistry, particularly CO as it is the main opacity source from $\sim$4.5 - 5.5 $\upmu$m for early L dwarfs \citep{Adams2023} and H$_{3}$O$^{+}$ from $\sim$7.5 - 8.5 $\upmu$m.
H$_{3}$O$^{+}$ has been postulated as a potential auroral tracer, due to H$_{3}^{+}$, a known auroral product in our solar system, reacting with H$_{2}$O \citep[]{Helling_2014,Gibbs2022Spectrograph,Pineda2024ImpactEmission}. H$_{3}$O$^{+}$ has a longer dynamical timescale than H$_{3}^{+}$, which may lead it to being observable \citep{Helling_h3o_2019}. These wavelengths are accessible with the JWST, in particular the NIRSPec G395H and MIRI MRS instruments. High resolution observations are more sensitive to detecting auroral emission in the spectrum, as shown in \cite{Faherty2024MethaneDwarf}.

\subsection{Comparison to previous observations: long term stability}

L/T transition objects are typically variable on rotational as well as longer-term timescales \citep[]{Apai_2013,Zhou2022RoaringAtmosphere,Biller_2024, Fuda2024,Chen2025}. Their light curves have been observed to evolve significantly from one rotation to another. However, this is not necessarily the case at all spectral types.
There are few examples of variability remaining stable on longer timescales.
{These largely stem from the Spitzer variability surveys of \cite{metchev2015weather} and \cite{Vos2022spin}. These include the objects, 2M0030-14, 2M0031+57, 2M0718-64, 2M0809+44 and 2M2228-43, which span the spectral types L6-T6. 2M1047+21, a T6.5 dwarf with an infrared period of 1.74 hours also displayed stable variability for over 7 rotations with Spitzer \cite{Allers2020}. These objects all displayed variability that did not evolve significantly over an observation duration of at least five rotations each. These objects all have infrared periods  between 1.08-1.64 hours, except for the L6 dwarf 2M0030-14, which has a period of 4.22$\pm$0.02 hours. These objects typically rotate more rapidly than the three in our sample. }

Long term stability in the light curve structure suggests that long lived features are responsible for the variability \citep[]{gizis2015kepler, croll2016long}.
In order to assess whether our objects show long term light curve evolution or display long term stability, we compared our results to light curves published in the literature. In our unique case of two sets of observations for 2M0036+18, we compare the two epochs directly. For 2M1721+33 we compare our observations with Spitzer 3.6~$\upmu$m observations from \cite{metchev2015weather}. For 2M1906+40, we {discuss the similarity between} our observations {and the} Kepler light curves from \cite{Gizis2013}.

We analysed our unique case of two HST observations of 2M0036+18. Despite both observations being just over 16 months apart (Table \ref{tab: observations}), both the light curves (Figure \ref{fig:LCs+period}) and spectral variability amplitude (Figure \ref{fig:cloud_mode+spot_models}) appear very similar. The observations cover slightly different phases of rotation, as can be seen in {the phase folded light curves of Figure \ref{fig: 0036_phase_fold}.} The spectral variability amplitude of both observations are very similar too, as shown in Figure \ref{fig:cloud_mode+spot_models}. The second observation has a slightly higher spectral variability amplitude from $1.1 - 1.25 ~ \upmu$m, but this may be attributed to the fact that this epoch probes the peak of the {ligth curve} amplitude more than the first epoch. Due to gaps in our observations caused HST's low Earth orbit, we did not achieve full phase coverage with either observation of 2M0036+18. Therefore, when calculating the spectral variability amplitude, the brightest and dimmest 5\% of spectra originate from different phases of 2M0036+18 in our two epochs.
The light curve amplitude of both observations are also consistent with each other within uncertainties, $1.41^{+0.16}_{-0.02}$ \% and $1.36^{+0.11}_{-0.01}$ \% for the first and second epochs respectively. \cite{croll2016long} also found that the light curves of 2M0036+18 were stable across observations spanning 122 days.

2M1721+33 was previously observed with Spitzer using the $3.6 ~ \upmu$m filter \citep{metchev2015weather}. We show the $3.6 ~ \upmu$m light curve in Figure \ref{fig:Spitzer LC}, along with the \cite{metchev2015weather} fit to the light curve and our {\tt celerité2} Gaussian process fit for this Spitzer light curve. As discussed in Section \ref{sec: light curve fitting}, the 2M1721+33 light curve contains two troughs, one deeper than the other. This shape is consistent across both epochs and wavelengths, however, the amplitudes differ. The Spitzer $3.6 ~\upmu $m light curve has an amplitude of $0.33 \pm0.07 ~\%$, while the HST light curve has an amplitude of $0.53^{+0.03}_{-0.01} ~\%$. {Yet}, we do not expect the amplitude {or the light curve shape} to be the same at different wavelengths as they probe different depths in the atmosphere \citep{Vos2020}.
The Spitzer observations were taken in 2011, nine years before the HST observations from this program. Much like 2M1906+40, this consistency in the light curve nine years apart suggests long-lived surface features. In this case, potentially two features causing a double trough shape in the light curve. {We note that since these observations were taken at different wavelengths, they probe different pressure levels \citep[{e.g.}][]{buenzli2012, Buenzli2014,McCarthy2025_JWST_SIMP}. However, it is interesting to note the shared similarities across both time and wavelength in the light curve shape.}

Although our observations of 2M1906+40 covered less than one rotation, it was observed by the Kepler Space Telescope for 887 days \citep[]{Gizis2013, gizis2015kepler}. The Kepler light curve showed a sinusoidal shape over the entire observation duration. This shape is consistent across both the Kepler and HST observations, taking place from 2011-2013 and 2020 respectively.
2M1906+40 was also observed with the Spitzer 3.6~$\upmu$m and 4.5~$\upmu$m channels for 20 hours on 17-18 October 2013 \citep{gizis2015kepler}. This study found that the 3.6~$\upmu$m light curve also exhibits stable sinusoidal variability, although variability was not detected at 4.5~$\upmu$m.
The amplitude of the light curves in each wavelength range are different however. The optical \textit{Kepler} light curve had an amplitude of 1.5~\% \citep{gizis2015kepler} the HST light curve from this work has an amplitude of $ 0.64^{+0.03}_{-0.02} ~\%$ and the Spitzer 3.6~$\upmu$m  light curve has an amplitude of 1.1~\%. {As with 2M1721+33,} we must note that none of these amplitudes were measured at the same epoch, {nor probe the same wavelength range, yet all share a common light curve shape.}

The fact that all three of our objects display {stable} light curve{s} across many years implies that there may be a different regime of variability in the early L-dwarfs.
L/T transition objects such as SIMP 0136, VHS 1256b and WISE 1049AB have variability that evolves from one rotation to the next \citep[]{ Zhou2022RoaringAtmosphere, Biller_2024, McCarthy2025_JWST_SIMP}. These objects have been observed across multiple epochs with both HST and JWST and they each show changes in their overall light curve amplitude as well as the shapes of their light curves. This light curve evolution may be due to dynamical evolution of clouds, carbon chemistry or hotspots on their surfaces \citep{McCarthy2025_JWST_SIMP}.
In contrast, the early L-dwarfs presented in this work appear somewhat stable over timescales of years. 

As all three of our objects have shown evidence for little light curve evolution over many years, this suggests that long-lived features may be present on all three objects. Our results show that {the variability} is likely caused by {clouds of different properties} or a magnetic spot. 
Future observations with telescopes such as JWST, across both the near and mid-infrared, would provide insight into whether clouds, magnetic spots and aurorae drive variability across the full infrared spectrum. This would allow for light curve comparison across multiple epochs in order to further analyse the light curve evolution of these objects. An observation with NIRSpec PRISM, for example, would obtain a light curve from 1-5 $\upmu$m and provide comparisons with the HST light curves from 1.1-1.67 $\upmu$m as well as the Spitzer 3.6 $\upmu$m and 4.5 $\upmu$m light curves.

\subsection{Implications for directly imaged exoplanets}

{While spectroscopic variability observations can be conducted on a large number of free floating brown dwarfs \citep[{e.g.}][]{Buenzli2014,Radigan2014,metchev2015weather,Vos2022spin}, variability studies of directly imaged exoplanets have been limited by the presence of a bright host star.
\cite{Apai2016} and \cite{Biller2021} searched for variability in the HR8799 system of exoplanets with the Very Large Telescope's SPHERE instrument, however neither study was able to confidently detect variability. This illustrates the challenge of variability observations from even the best ground based telescopes. There have been attempts to observe variability with space based telescopes too. \cite{Zhou2019PMcompanion} observed three planetary mass companions with HST, but only found tentative detections of photometric variability. \cite{Zhou_2020_high_contrast} observed the directly imaged exoplanet HD 106906b with HST in the F127M, F139M and F153M filters. Tentative variability was detected with in the F127M light curve but the other two filters detected no variability. These results illustrate the difficulty of obtaining photometric variability of directly imaged exoplanets.}

{Recent observations with JWST have revealed variability in the directly imaged planetary mass companion 2M1207 b  \cite{Adams2025_2M1207}. Other ongoing programs, such as \cite{Yifan_beta_pic_b_proposal}, will search for variability in directly imaged exoplanets such as $\beta$ pic b with JWST.}

Brown dwarfs can be considered as analogues for directly imaged exoplanets.
There are a number of directly imaged exoplanets that share similar temperatures and spectral types to the early L-dwarfs presented in this paper, such as $\beta$ pic b, RXS 1609 b and AB Pic b \citep[]{lagrange2009, Bowler2013, Palmabifani2023}. At these temperatures the effects of clouds, aurorae and magnetic spots may all be important. By studying the variability of early L-dwarfs, we can establish the dominant processes in extrasolar atmospheres at these temperatures without the need for high-contrast imaging. Early L-dwarfs such as 2M1721+33, 2M0036+18 and 2M1906+40 are analogues for directly imaged exoplanets such as those listed above. 

Our results indicate that early L-dwarfs are variable and exhibit high variability amplitudes. Clouds and magnetic spots are the most plausible variability mechanisms for early L-dwarfs from 1.1-1.67$~\upmu$m.
Therefore, it is likely that directly imaged exoplanets with similar effective temperatures and spectral types are variable. Our work shows that this variability is likely driven by condensate clouds and/or magnetic spots at near-IR  wavelengths (1.1-1.67$~\upmu$m), but auroral variability may become important at longer wavelengths. {Our spectroscopic variability results help put photometric variability of directly imaged exoplanets into context.} Future high contrast variability searches with JWST, VLT and ELT are likely to reveal variability in directly imaged exoplanets with early L spectral types.

\section{Conclusions}\label{sec:conclusions}

In this work, we present HST/WFC3 variability observations of three early L-dwarfs, 2M1721+33, 2M0036+18 and 2M1906+40. We use a modelling framework which combines self-consistent forward models and atmospheric retrievals to model the spectral variability amplitude of each object for three potential cases: cloud-driven variability, auroral variability and magnetic spot-driven variability. We were able to assess the likely drivers of variability in each object. 

\begin{enumerate}

    \item All three brown dwarfs show significant variability across all wavelengths, with with white light curve amplitudes of 
    $0.64^{+0.03}_{-0.02}$~\%, $1.41^{+0.16}_{-0.02}$ \%, (obs 1), 
    $1.36^{+0.11}_{-0.01}$ \% (obs 2), and
    $0.53^{+0.03}_{-0.01}$ \%,
    for 2M1721+33, 2M0036+18 and 2M1906+40 respectively. 

    \item We report a new period of $4.93^{+0.36}_{-0.22}$ hours for 2M1721+33 from our Gaussian process fit to our new HST and archival Spitzer light curves. This resulted in this program not capturing a fully phase resolved light curve. Our measured periods for 2M0036+18 and 2M1906+40 agree with the literature values.

    \item The colour modulation of our objects extend the trend discussed in \citet{Lew_colour}, where early L-dwarfs become brighter and bluer as they rotate. However, this changes for the L1 target 2M1906+40 which has near grey modulations.

    \item We have developed a modelling framework by combining self consistent forward models and atmospheric retrievals to reproduce the spectral variability amplitude of brown dwarfs for variability that is driven by changing clouds, aurorae and magnetic spots. In the future, this can be applied to other objects with observations at different wavelengths to model their spectral variability amplitude.

    \item Our analysis has found that either clouds or magnetic spots are plausible drivers of variability in the early L dwarfs 2M1721+33, 2M0036+18 and 2M1906+40. Aurora can be ruled out as a driver of variability at the measured wavelengths of 1.1-1.67~$\upmu$m. However, future observations at longer wavelengths are likely more sensitive to potential auroral variability, as they probe shallower atmospheric depths.

    \item All three of our objects have shown evidence for long-term stability of their light curves.  In the case of long-term light curve stability, the variability is likely driven by a long-lived feature. This may be a cloudy spot {or region}, or a stellar-like magnetic spot. Future observations across the near-infrared and mid-infrared are required in order to disentangle this further.

\end{enumerate}

This work has presented a method to disentangle the drivers of variability in early L-dwarf atmospheres based on their spectral variability behaviour. Future studies with instruments such as NIRSpec G395H and MIRI MRS onboard JWST would allow us to further characterise these processes. At longer wavelengths, we can search for variability in the 10$~\upmu$m silicate feature to disentangle whether the long-lived features on these objects are cloudy or magnetic in nature. Searches for aurora on these worlds would benefit from observations that probe higher in the atmosphere, at longer wavelengths, from 4.5-5.5 $\upmu$m and 7.5 - 8.5 $\upmu$m, to detect species that may be affected by auroral heating. This work has provided a flexible modelling framework that can be applied to any L, T or Y dwarf to determine what the likely drivers of variability are across different wavelengths and resolutions. This provides an insight into the likely variable nature of directly imaged exoplanets that will be observed with next generation telescopes.

\begin{acknowledgements}
    We would like to thank the referee for their helpful comments which helped improve the quality of this paper. We would like to thank Ben Lew for their guidance on fitting the colour modulation using the ODR method. We would like to thank Caroline Morley for their guidance with the Sonora Diamondback contribution function.
    CO’T, JMV and EN acknowledge support from a Royal Society - Research Ireland University Research Fellowship (URF/1/221932, RF/ERE/221108). MS acknowledges support from Trinity College Dublin via a Trinity Research Doctoral Award. This research is based on observations made with the NASA/ESA Hubble Space Telescope obtained from the Space Telescope Science Institute, which is operated by the Association of Universities for Research in Astronomy, Inc., under NASA contract NAS 5–26555. These observations are associated with program 15924.
\end{acknowledgements}

\bibliographystyle{yahapj}
\bibliography{references}

\appendix

\onecolumn

\section{Atmospheric retrieval corner plots}\label{app: Appendix_big_corners}

This section of the Appendix displays the corner plots  for each of the petitRADTRANS atmospheric retrievals for each observation.

\begin{figure*}[ht!]
    \centering
    \includegraphics[width= \textwidth]{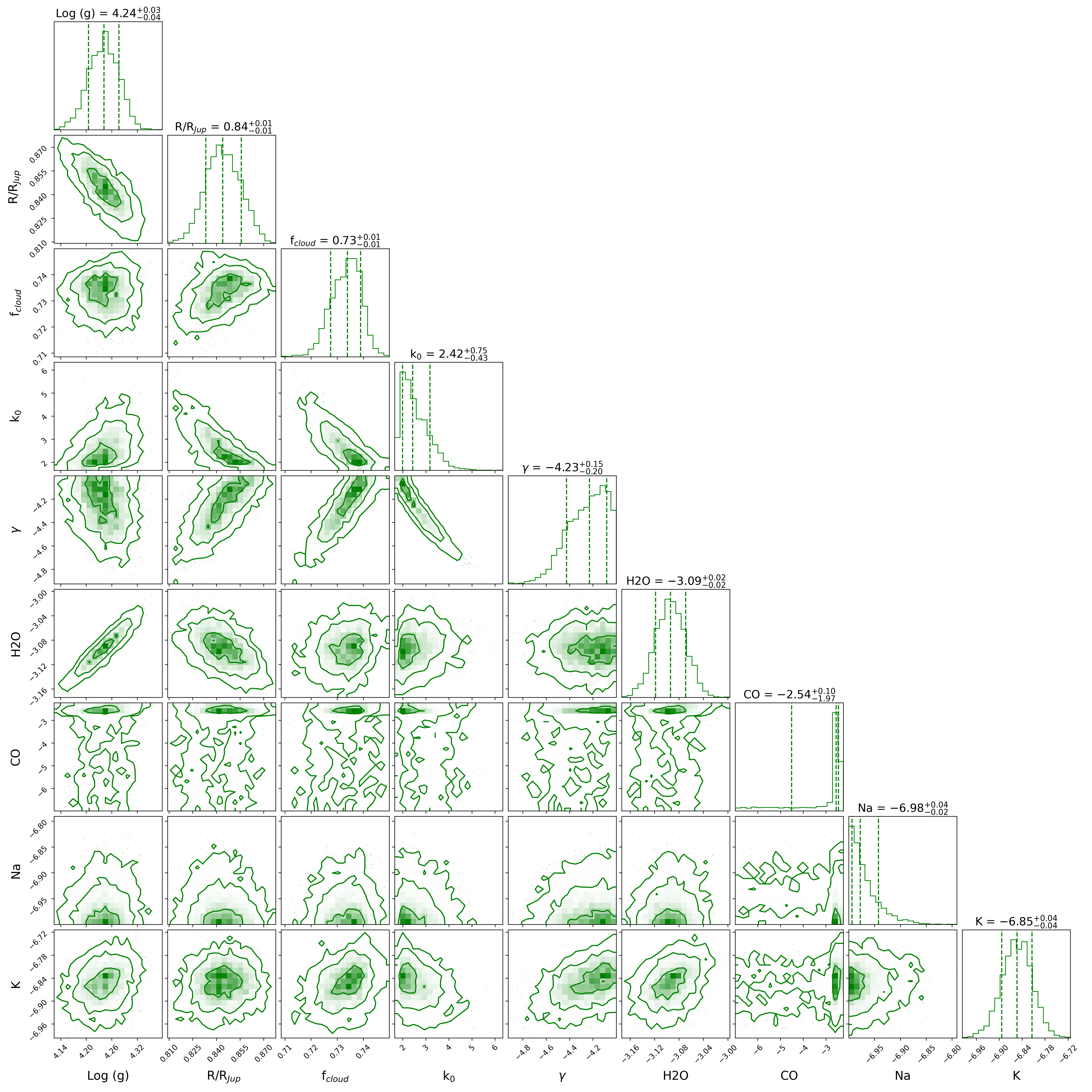}
    \caption{Posterior distributions for the petitRADTRANS atmospheric retrieval of 2M1721+33.  }
    \label{fig: 1721_ret_corner}
\end{figure*}

\begin{figure*}[ht!]
    \centering
    \includegraphics[width= \textwidth]{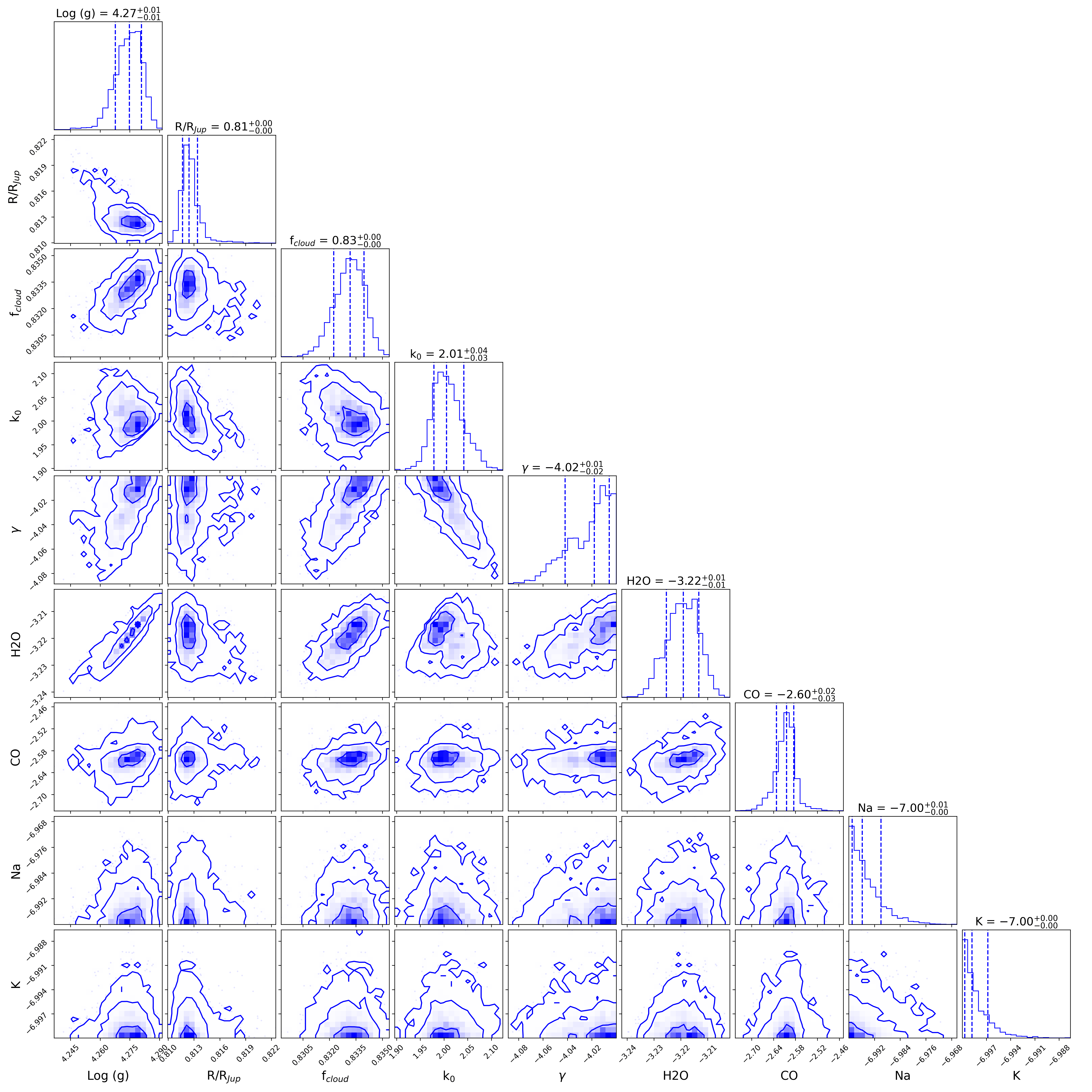}

    \caption{Posterior distributions for the petitRADTRANS atmospheric retrieval of 2M0036+18 (epoch 1).}
    \label{fig: 0036_OBS1_ret_corner}
\end{figure*}

\begin{figure*}[ht!]
    \centering
    \includegraphics[width= \textwidth]{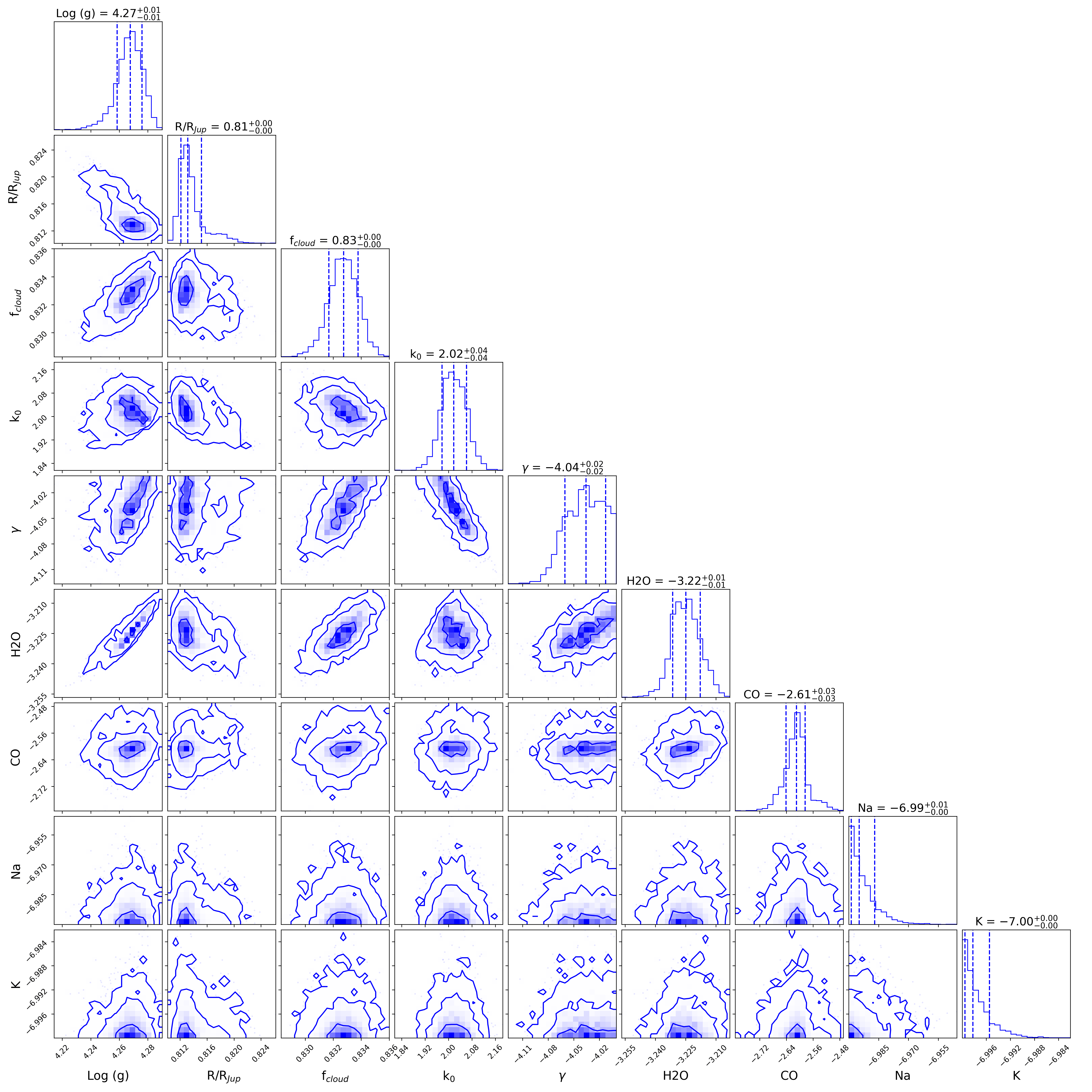}
    \caption{Posterior distributions for the atmospheric retrieval of 2M0036+18 (epoch 2).}
    \label{fig: 0036_OBS2_ret_corner}
\end{figure*}

\begin{figure*}[ht!]
    \centering
    \includegraphics[width= \textwidth]{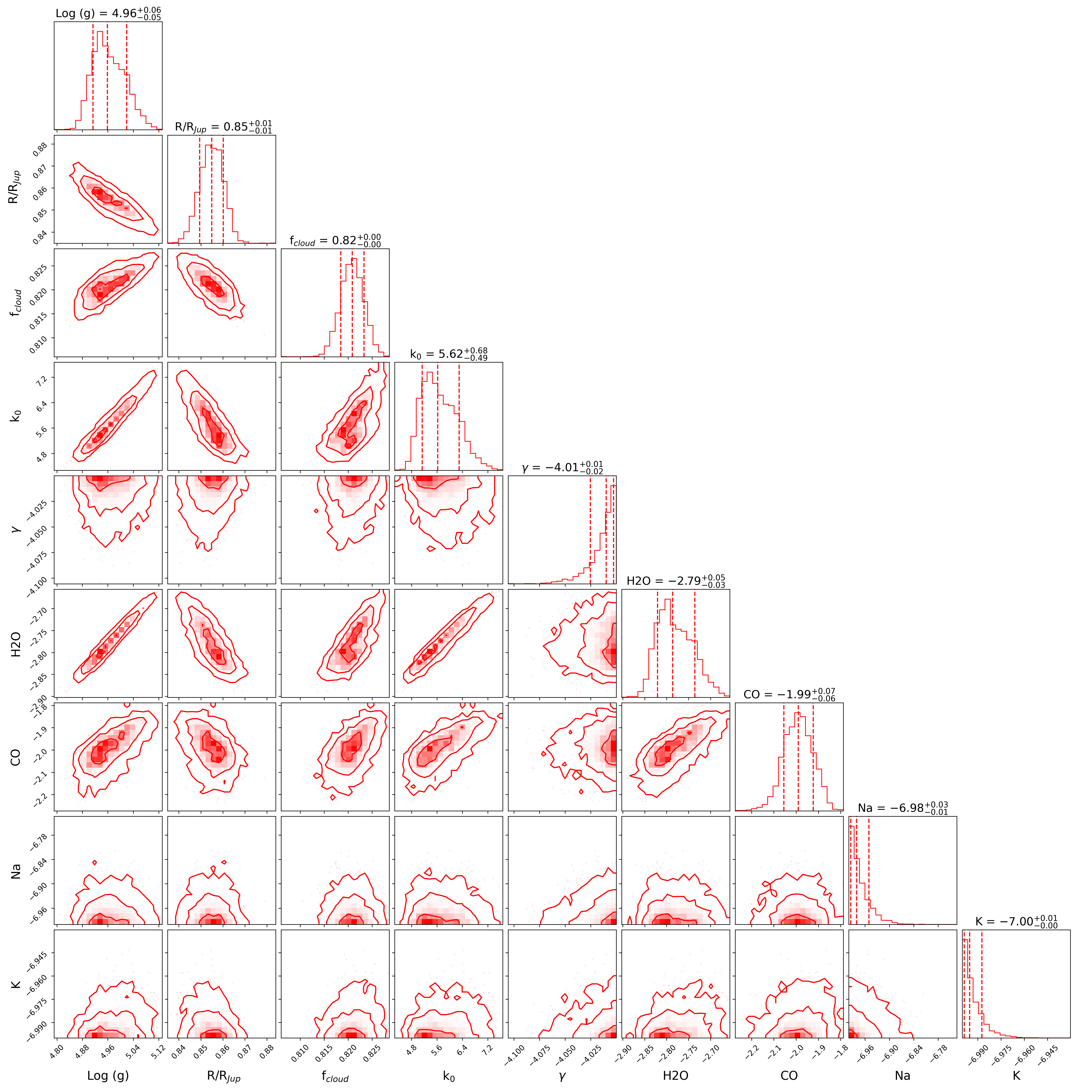}
    \caption{Posterior distributions for the petitRADTRANS atmospheric retrieval of 2M1906+40.  }
    \label{fig: 1906_ret_corner}
\end{figure*}

\clearpage
\twocolumn[
\section{Cloud variability model corner plots}\label{app: Appendix_Cloud_corners}
This section of the Appendix displays the corner plots for the cloud driven variability models. We show the posteriors for the $\Delta$A, $\gamma$ and k$_{0}$  values for each observation.]

\begin{figure}[ht!]
    \centering
    \includegraphics[width= \columnwidth]{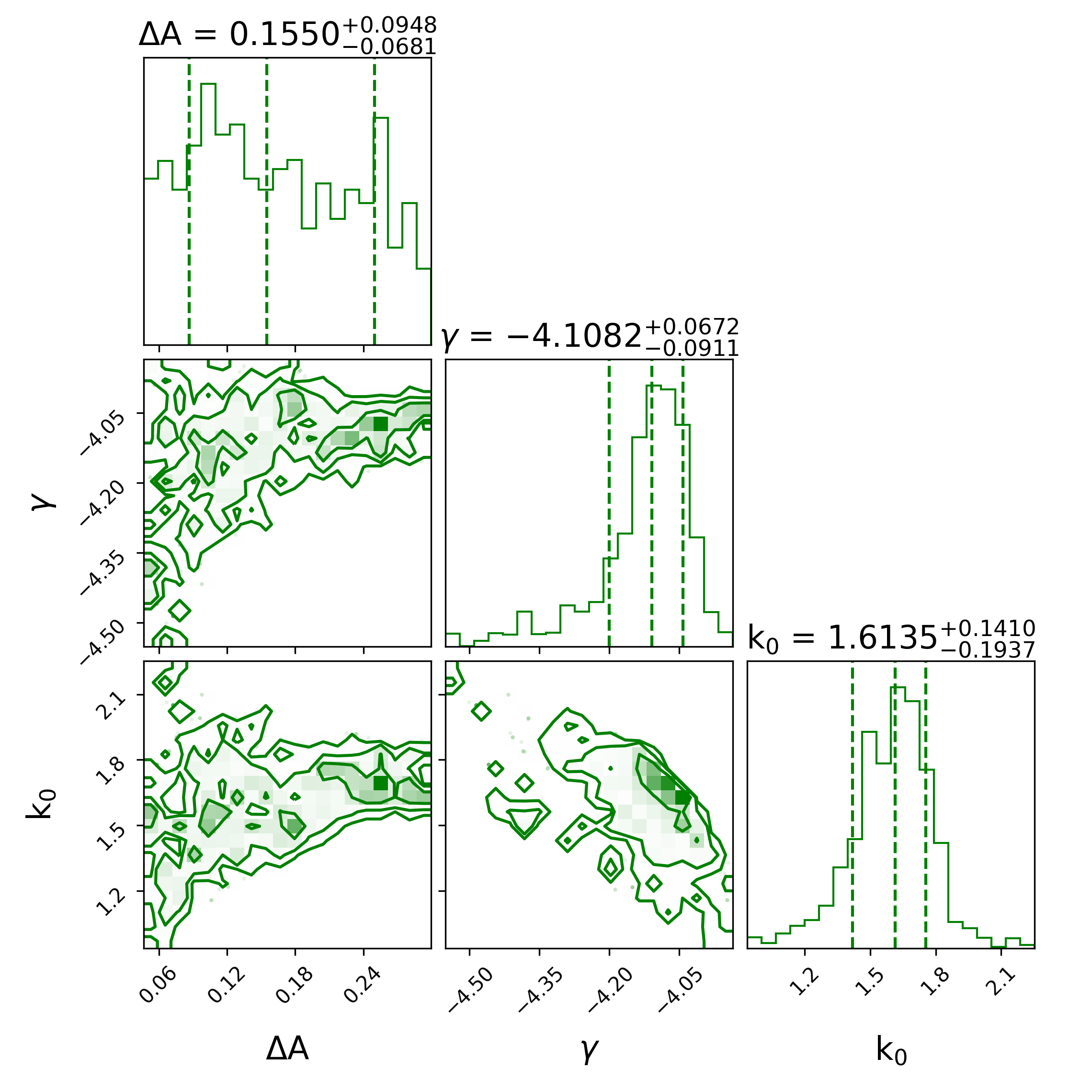}
    \caption{Posterior distributions of the cloud model parameters $\Delta$A, $\gamma$ and k$_{0}$ for 2M1721+33.  }
    \label{fig: 1721_corner}
\end{figure}

\begin{figure}[ht!]
    \centering
    \includegraphics[width= \columnwidth]{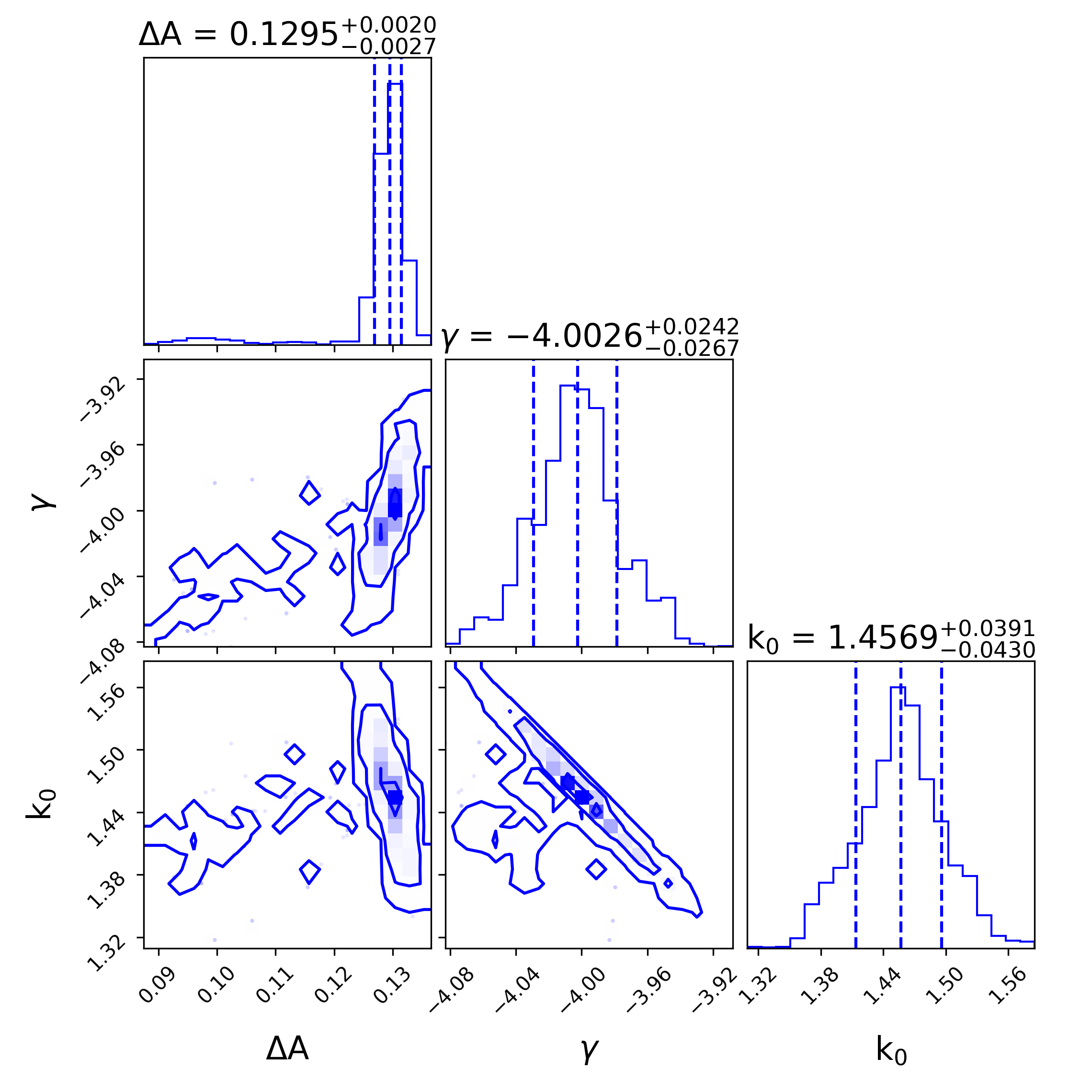}
    \caption{Posterior distributions of the cloud model parameters $\Delta$A, $\gamma$ and k$_{0}$ for 2M0036+18 (epoch 1).}
    \label{fig: 0036_obs1_corner}
\end{figure}

\begin{figure}[ht!]
    \centering
    \includegraphics[width= \columnwidth]{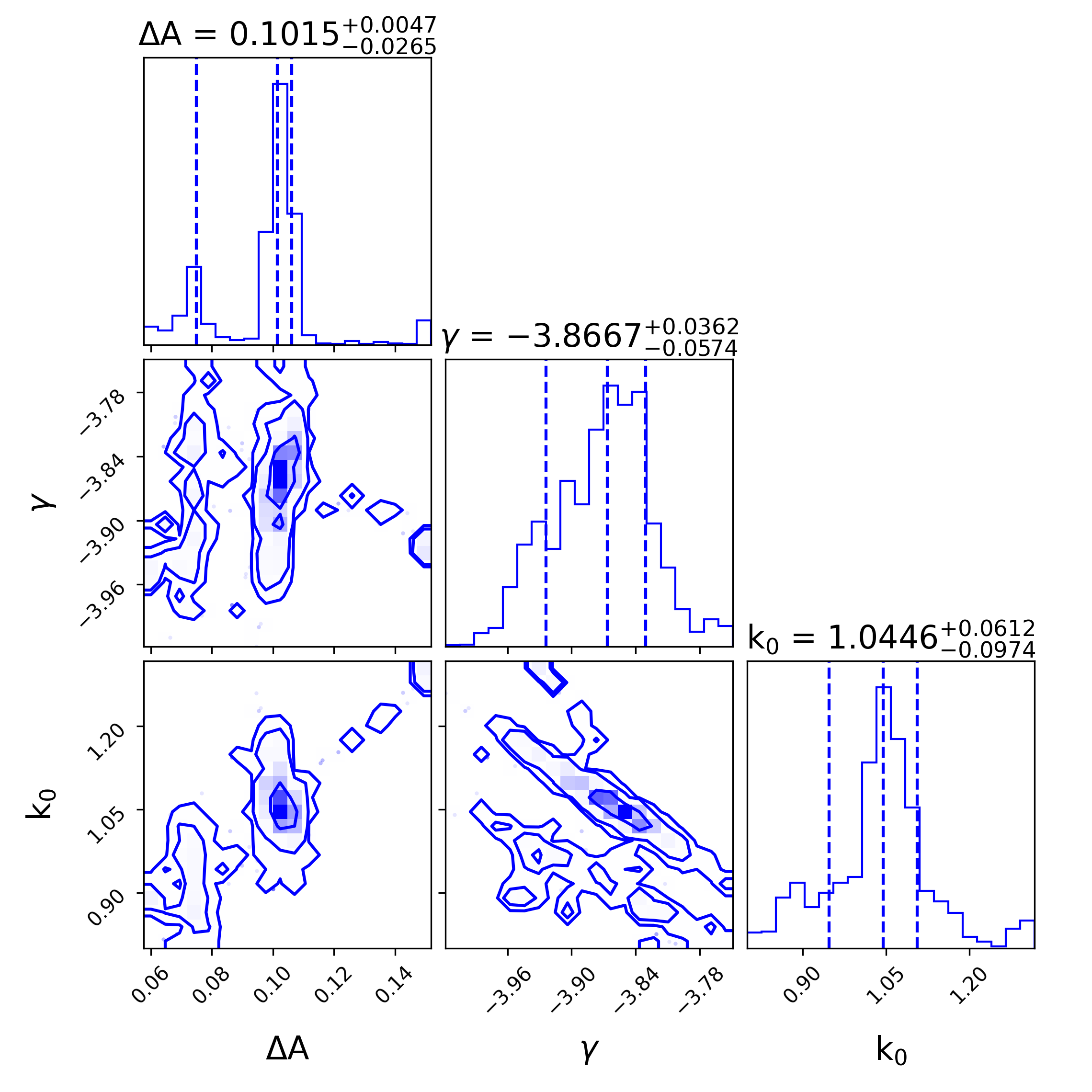}
    \caption{Posterior distributions of the cloud model parameters $\Delta$A, $\gamma$ and k$_{0}$ for  2M0036+18 (epoch 2).}
    \label{fig: 0036_obs2_corner}
\end{figure}

\begin{figure}[ht!]
    \centering
    \includegraphics[width= \columnwidth]{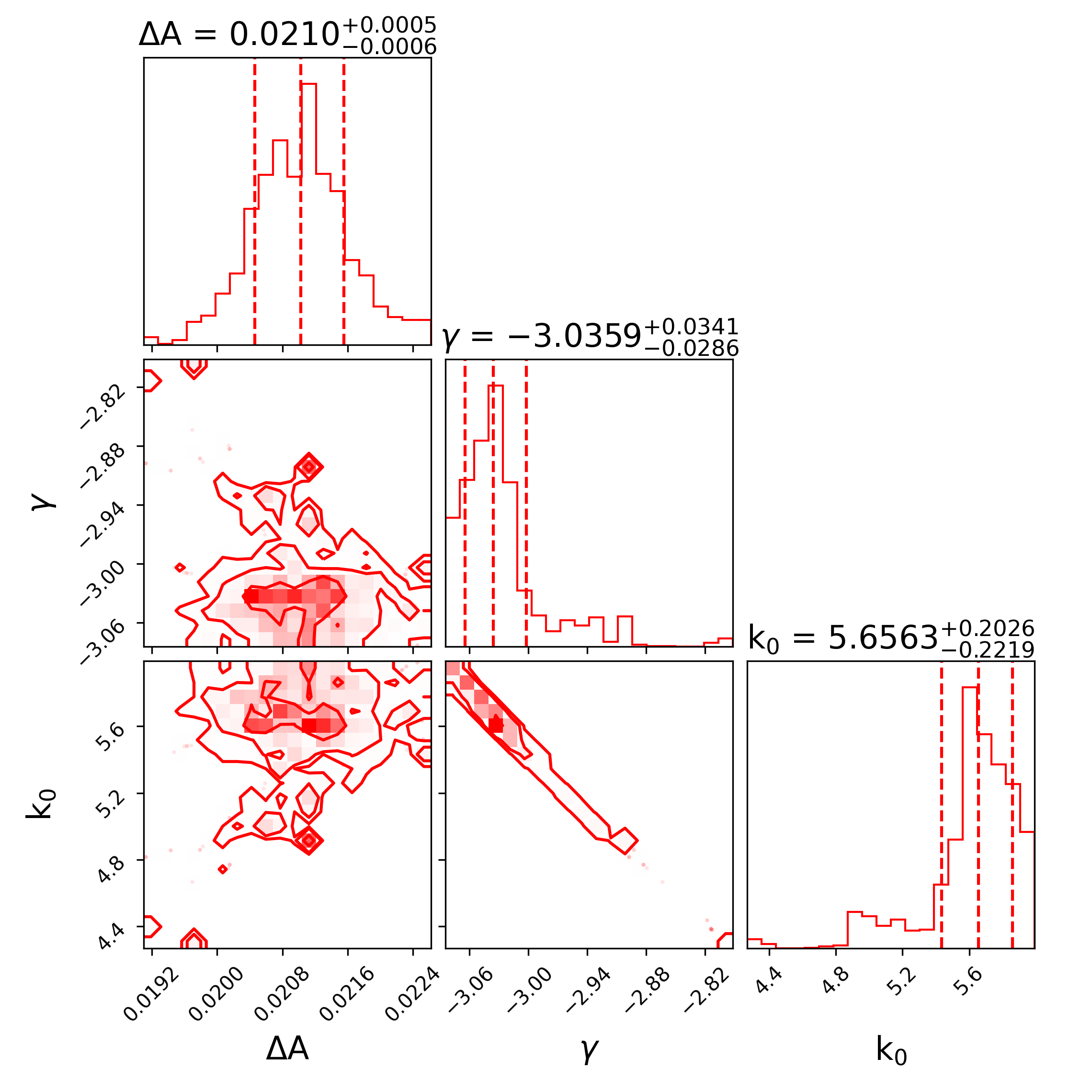}
    \caption{Posterior distributions of the cloud model parameters $\Delta$A, $\gamma$ and k$_{0}$ for 2M1906+40.}
    \label{fig: 1906_corner}
\end{figure}

\newpage
\onecolumn
\section{Magnetic spot variability posterior distribution Plots}\label{app: Appendix_Mag_corners}

This section of the Appendix displays the posterior distributions for the spot driven variability models. We show the posteriors for the $\Delta$A value for each observation for the best fit $\Delta$T.

\begin{figure*}[h]
    \centering
    \includegraphics[width= \textwidth]{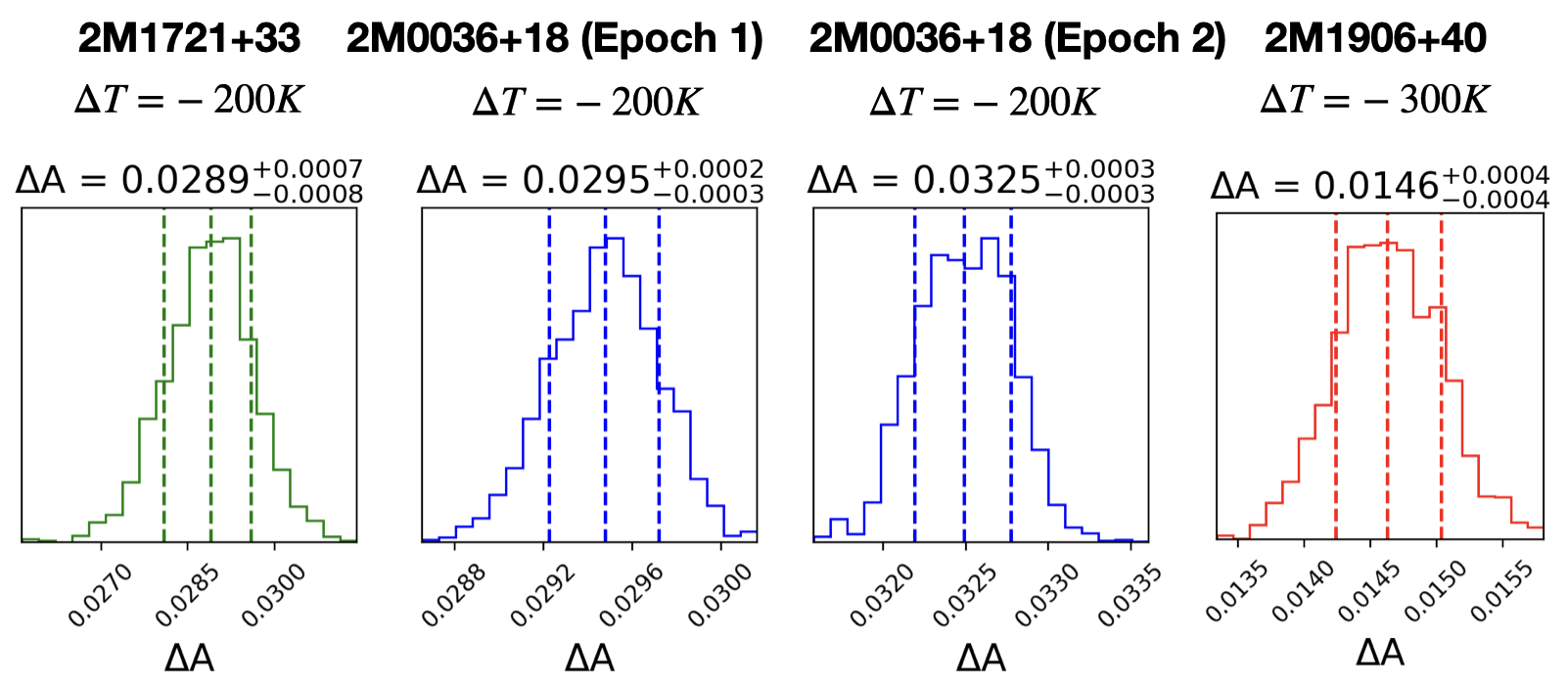}
    \caption{Posterior distributions of the best fit magnetic spot model, $\Delta$A parameter, for each object. }
    \label{fig: spot_mcmc}
\end{figure*}

\end{document}